\documentclass[aps,prb,twocolumn,superscriptaddress,preprintnumbers]{revtex4}
\usepackage[colorlinks,bookmarks=false,citecolor=blue,linkcolor=red,urlcolor=blue]{hyperref}
\usepackage{epsfig}
\usepackage{tikz}
\usepackage{graphicx}

\usepackage{dcolumn}
\usepackage{bm}

\usepackage{amsmath,amssymb}
\usepackage{color}

\makeatletter
\renewcommand*\env@matrix[1][*\c@MaxMatrixCols c]{%
\hskip -\arraycolsep
\let\@ifnextchar\new@ifnextchar
\array{#1}}
\makeatother

\usepackage{appendix}

\newcommand{\vS}{\vec{S}}

\newcommand\bea{\begin{eqnarray}}
\newcommand\eea{\end{eqnarray}}
\newcommand\beq{\begin{equation}}
\newcommand\eeq{\end{equation}}

\newcommand{\noi}{\noindent}
\newcommand{\non}{\nonumber}
\newcommand{\al}{\alpha}
\newcommand{\de}{\delta}
\newcommand{\De}{\Delta}

\newcommand{\lm}{\lambda}

\newcommand{\om}{\omega}
\newcommand{\da}{\dagger}

\newcommand{\la}{\langle}
\newcommand{\ra}{\rangle}

\begin{document}

\title{Effective theories for quantum spin clusters: Geometric phases and state selection by singularity}
\author{Subhankar Khatua}
\email{subhankark@imsc.res.in}
\affiliation{The Institute of Mathematical Sciences, HBNI, C I T Campus, 
Chennai 600113, India}
\author{Diptiman Sen}
\email{diptiman@iisc.ac.in}
\affiliation{Centre for High Energy Physics, Indian Institute of Science, 
Bengaluru 560012, India}
\author{R. Ganesh}
\email{ganesh@imsc.res.in}
\affiliation{The Institute of Mathematical Sciences, HBNI, C I T Campus, 
Chennai 600113, India}
\date{\today}

\begin{abstract}
Magnetic systems with frustration often have large classical degeneracy. 
We show that their low-energy physics can be understood as dynamics within the space of classical ground states. We demonstrate this mapping in 
a family of quantum spin clusters 
where every pair of spins is connected by an $XY$ antiferromagnetic bond. 
The dimer with two spin-$S$ spins provides the simplest example -- 
it maps to a quantum particle on a ring ($S^1$). The trimer is more complex, 
equivalent to a particle that lives on two disjoint rings ($S^1\otimes \mathbb{Z}_2$). It has an additional subtlety for half-integer $S$ values, wherein both rings must be threaded by $\pi$-fluxes to obtain a satisfactory mapping. This is a consequence of the geometric phase incurred by spins. For both the dimer and the trimer, the validity of the effective theory can be seen from a path-integral-based derivation.
This approach cannot be extended to the quadrumer which has a non-manifold ground state space,
consisting of three tori that touch pairwise along lines. 
In order to understand the dynamics of a particle in this space, we develop a tight-binding model with this connectivity. Remarkably, this successfully reproduces the low-energy spectrum of the quadrumer. For half-integer spins, a geometric phase emerges which can be mapped to two $\pi$-flux tubes that reside in the space between the tori.
The non-manifold character of the space leads to a remarkable effect -- the dynamics at low energies is not ergodic as the particle is localized around singular lines of the ground state space. 
The low-energy spectrum consists of an extensive number of bound states formed around singularities. 
Physically, this manifests as an order-by-disorder-like preference for collinear ground states. However, unlike order-by-disorder, this `order by singularity' persists even in the classical limit. We discuss consequences for field theoretic studies of magnets.
\end{abstract}
\pacs{}\keywords{}
\maketitle

\section{Introduction} 

A guiding principle in physics is to seek effective low-energy theories. 
Apart from describing the system at low temperatures, this can 
reveal `emergent' properties that may not resemble the gross system 
or its microscopic constituents. 
This approach has a long and successful history in magnetism. Examples include 
Haldane's field theory for spin chains~\cite{Haldane1983,Haldane1983b,Affleck1988}, 
spin ice physics~\cite{Savary2017}, Luttinger liquid 
theory~\cite{Luttinger1,Luttinger2,Luttinger3}, and so on.
To build such a theory for a macroscopic system (e.g., a three-dimensional magnet), an appropriate starting point is its smallest building block or motif. This is exemplified in triangle-motif-based Heisenberg antiferromagnets. A single triangle, at low energies, maps to a rigid rotor described by an $SO(3)$ 
matrix~\cite{Kawamura1984}. Starting from this insight, an $SO(3)$ field theory can be constructed to describe macroscopic magnets~\cite{Dombre1989,Rao1994}. 
In this article, we derive effective theories for a class of clusters/motifs with frustration. Even at the level of a single cluster, we find surprising results that suggest broadly applicable principles.

A characteristic feature of frustrated magnets is large classical 
degeneracy. Treating each spin as a classical vector, there are multiple ways to orient spins so as to minimize the energy -- the set of all such states is 
the classical ground state space (CGSS). Using this notion, we may state 
a general principle: \textit{the low-energy dynamics of a cluster of 
quantum spins is equivalent to the problem of a particle moving in the CGSS}. 
Heuristically, this equivalence is expected to hold in the semiclassical limit, i.e., when $S$, the spin quantum number, is large. Below, we examine this principle in clusters with increasing complexity. We find this principle to hold true in all cases, as long as $S$ is not too small. Two subtleties emerge from our 
analysis: (a) An appropriate Aharonov-Bohm flux must be threaded through the 
CGSS to incorporate Berry phase effects. (b)
If the CGSS forms a smooth manifold, the equivalence can be readily derived using a path integral approach. In some systems, the CGSS may have a non-manifold structure due to singularities. 
We empirically 
find that the principle still holds. Remarkably, such 
singularities can give rise to a localizing effect, which we call `order 
by singularity'.

This phenomenon shares similarities with 
the well-established notion of `order by disorder'\cite{Chalker2011}. The central idea here is that fluctuations can stabilize ordered phases\cite{Villain1980, Shender1982,Henley1989}. This plays a key role in frustrated systems which typically have large classical degeneracies that are `accidental', i.e., unrelated to symmetries of the Hamiltonian. In this work, we will focus on quantum fluctuations, regulated by $1/S$, that can break this degeneracy, e.g., by contributing differing zero-point contributions to the ground state energy. We do not expect such selection effects to survive in the classical limit ($S\rightarrow \infty$ ) where, by definition, all fluctuations are suppressed.

We re-express this notion of order by disorder as follows. In the limit of large-$S$, assuming zero temperature, a 
magnetic cluster samples every point on the CGSS. Equivalently, it maps to 
a particle whose low-energy states are uniformly supported in the CGSS. 
Weak quantum (finite $S$) fluctuations introduce a potential on the 
CGSS so that the particle is localized near the potential minima (see, for
instance, Ref.~\onlinecite{Henley1998}). This effect becomes progressively 
weaker as we approach the classical limit ($S\rightarrow\infty$).

We distinguish this from order by singularity, a stronger effect that persists even 
in the classical limit. The defining feature of the latter is the formation of low-energy bound states around singularities. The low-energy dynamics of a particle on the CGSS becomes non-ergodic, tied down to these bound states. Unlike order by disorder which is ubiquitous in frustrated magnets, 
order by singularity is a rare effect requiring the presence of singularities as a necessary condition. 

\begin{figure}
\includegraphics[width=\columnwidth]{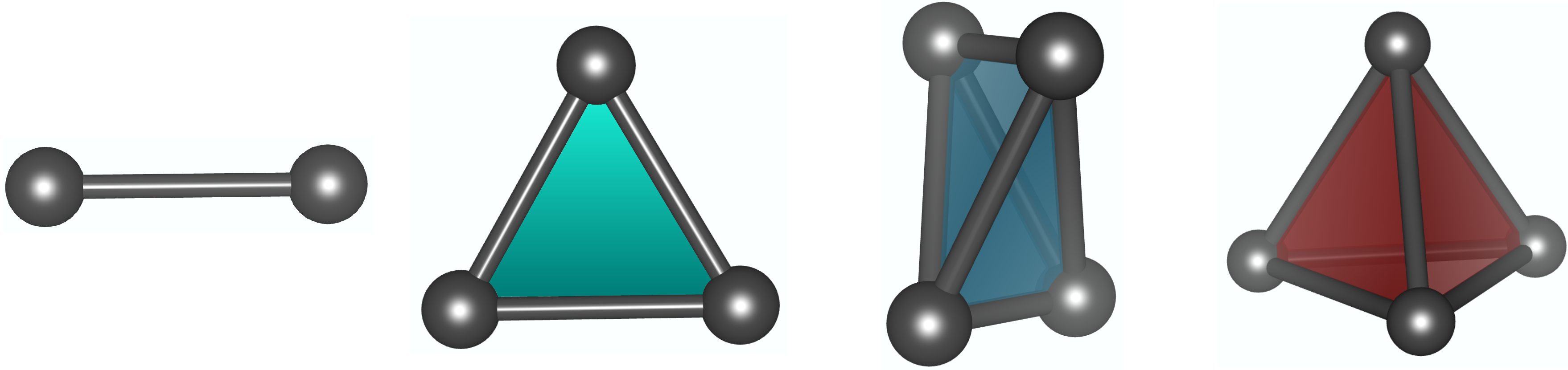}
\caption{Clusters studied here (from left to right): dimer, trimer, asymmetric quadrumer and quadrumer. In each case, every pair of spins is coupled by an $XY$ bond. The bond strengths respect the symmetries of the corresponding cluster. }
\label{fig.clusters}
\end{figure}

In this paper, we study the clusters shown in Fig.~\ref{fig.clusters}. The dimer, trimer and quadrumer consist of 2, 3 and 4 spins at the vertices of a rod, triangle and a tetrahedron respectively. They share the property that each 
pair of spins has the same spatial separation. Consequently, we assume that 
each pair of spins experiences the same coupling. We also study the asymmetric 
quadrumer which has reduced symmetry. We take all the bonds to be $XY$-like 
and antiferromagnetic, with the Hamiltonian
\begin{eqnarray} H ~=~ \sum_{i<j} ~J_{ij} \left[ {S}_i^x {S}_j^x ~+~ 
{S}_i^y {S}_j^y \right], \end{eqnarray}
where $i,j$ sum over $N$ spins, with $N=2,3$ and $4$. We set $\hbar = 1$ in all cases. In the dimer, trimer and quadrumer, the couplings are all equal, i.e., $J_{ij}=J>0$. In the asymmetric 
quadrumer, we have $J_{12} = J_{34} = (1+\lm) J$ and $J_{13}=J_{14}=
J_{23}=J_{24} = J$, with $J>0$ and $\lm>0$. 
The symmetric members of this family (dimer, trimer and quadrumer) are 
well-studied for the case of Heisenberg-like couplings. The Heisenberg dimer 
maps to a unit-vector field, or equivalently, a particle on a 
sphere~\cite{Haldane1983,Haldane1983b,Affleck1988}. The trimer maps to 
a rigid rotor, i.e., a particle in $SO(3)$ 
space~\cite{Kawamura1984,Dombre1989,Rao1994,Sen1993}. The quadrumer maps to 
a particle in a five-dimensional space with singular subspaces; this can be 
approximated as a rigid rotor and an emergent free spin~\cite{Khatua2018}.

The remainder of this article is structured as follows. In Sec.~\ref{sec.nlsm}, we derive the effective model for a system with an arbitrary CGSS manifold. This brings out the mapping to the picture of a particle moving on the CGSS.
In Sec.~\ref{sec.dimer}, we discuss the $XY$ dimer and show that its CGSS is a ring. We derive its effective low-energy theory and compare with the numerically 
obtained spectrum. In Sec.~\ref{sec.trimer}, we discuss the $XY$ trimer 
whose CGSS forms two disjoint rings. We present its effective theory which explains the numerically obtained spectrum. We consider the asymmetric quadrumer in Sec.~\ref{sec.asym_quad} which has a two-dimensional manifold as the CGSS. 
It provides a useful reference point for a discussion of the quadrumer and 
its non-manifold CGSS in Sec.~\ref{sec.quad}. In Secs.~\ref{ssec.quad_tb_int}-\ref{ssec.quad_qm_hint2}, we propose a tight-binding analog to explain the 
quadrumer spectrum, discussing the cases of integer and half-integer spins separately due to different Berry phase effects. Section~\ref{sec.ObDquad} discusses the role of order by disorder, showing that it is insufficient to explain the observed low-energy spectrum. Section~\ref{sec.ObSED} gives evidence for order by singularity from the spin model. In Sec.~\ref{sec.ObSTB}, we discuss order 
by singularity from the tight-binding point of view, demonstrating that the 
low-energy spectrum consists exclusively of bound states. We end with a 
summary and discussion in Sec.~\ref{sec.summary}.

\section{Effective low-energy theory for an arbitrary CGSS manifold}
\label{sec.nlsm}

The mapping between a magnet and a particle moving in the CGSS can be seen as follows. 
We use the well-known semiclassical path integral formulation for spin 
systems~\cite{Auerbach,Fradkin}. In this scheme, the path integral is over all trajectories of classical spin vectors of length $S$. The action, written as an expansion in $1/S$, consists of a Berry phase term and an energy term. The former can be given a geometric interpretation 
as the area swept out by each spin on the sphere.
The latter, at leading order, is simply the classical energy. For large $S$, paths within the CGSS dominate the path integral. Low-energy excitations can be taken into account as small fluctuations out of this space, taking the form of $1/S$ corrections. 
This paradigm can provide physical insight into the nature of the low-energy spectrum, e.g., the stationary states of a single spin with easy axis anisotropy are analogous to a particle tunnelling between two potential wells\cite{Loss1992}.
We provide a generic derivation here that is applicable to systems wherein the CGSS is a smooth manifold. We apply it to specific cases in Secs.~\ref{sec.dimer}, \ref{sec.trimer} and \ref{sec.asym_quad} below. We note that the arguments here do not extend to the case of the symmetric quadrumer, discussed in Sec.~\ref{sec.quad}.
 
Consider a zero-dimensional system (a cluster) with $N$ spins. This corresponds to a $(2N)$-dimensional classical configuration space, as each spin can be 
described by two variables (namely, polar and azimuthal angles). We assume a $d$-dimensional CGSS, described by coordinates $p_i$, where $i=1,\cdots,d$; we will
assume that the CGSS is a
$d$-dimensional manifold, where the $p$-coordinates can be defined in a smooth manner. At any point on the CGSS, we have `hard' fluctuations that cost energy, given by $q_l$, where $l=1,\cdots,2N-d$. The spins take the form
\begin{eqnarray} \vec{S}_k ~=~ S \frac{ \hat{n}_k (p_1,\cdots,p_d) + 
\vec{m}(q_1,\cdots,q_{2N-d})/S}{\sqrt{1+ \vec{m}_k\cdot \vec{m}_k/S^2 }}.
\label{eq.spinnlsm} \end{eqnarray}
Here, $k=1,\cdots,N$ labels the spins. We have introduced $\hat{n}_k$, a unit vector for each $k$. It orients spins so as to give rise to the ground state specified by the $p$ coordinates. The vector $\vec{m}_k$, determined by $q$ coordinates, introduces a deviation from the ground state space. In order to preserve normalization, we must have $\hat{n}_k\cdot\vec{m}_k=0$. This fixes the length of the spin to $S+\mathcal{O}(1/S)$. This definition is suitable for low energies and large $S$ values, where each spin has length $S$, but with $\mathcal{O}(1)$ fluctuations out of the ground state space. 

Parametrizing spins using Eq.~\eqref{eq.spinnlsm}, the leading order 
energy term in the action generically takes the form,
\begin{eqnarray} \int_0^\beta d\tau E(p,q) ~=~ \beta E_{cl} ~+~ \int_0^\beta 
d\tau ~\sum_{l,m=1}^{2N-d} A_{lm} {q}_l {q}_m,
\label{eq.ENLSM} \end{eqnarray}
where $E_{cl}$ is the classical ground state energy.
This can be deduced as follows. We first note that linear terms are not 
allowed due to the extremum nature of the classical ground states. Further, there can be no explicit dependence on the $p_i$'s as all points in the CGSS are degenerate. In the spirit of a low-energy theory, we consider the $q_l$'s to be small, keeping only quadratic terms. This can alternatively be seen as an expansion in $1/S$, keeping $\mathcal{O}(S^0)$ terms. The coefficients $A_{lm}$ can be determined for any specific case, as discussed in the following sections.

The Berry phase term in the action takes the form
\begin{eqnarray}
\non \mathcal{S_B} &=& iS\int_0^\beta d\tau ~\sum_{i=1}^N \vec{A}
(\hat{\Omega}_i)\cdot\partial_\tau \hat{\Omega}_i \\
&=&i 2\pi S Q ~+~ \int_0^\beta d\tau ~\sum_{i=1}^d \sum_{l=1}^{2N-d} B_{il} 
\dot{p}_i{q}_l. \label{eq.BerryNLSM} \end{eqnarray}
Here, $\vec{A}$ is the vector potential of a unit monopole charge at the center of a unit sphere, while $\hat{\Omega}_i (\tau)$ is a unit vector oriented along the $i^{\mathrm{th}}$ spin at time $\tau$~\cite{Auerbach,Fradkin}. 
We have a quantized contribution, $i2\pi S Q$, where $Q$ is an integer. 
This arises from trajectories within the ground state space, when spins sweep out non-zero areas in a closed loop. Its quantized nature arises from the planar
nature of the ground states in the systems studied here (due to the $XY$ nature of couplings). Moving along a loop within the
ground state space, each spin can move around the equator an integer number of times. Each pass covers an area corresponding to one hemisphere, $2\pi$. 
The sum of contributions for all $N$ spins has the form $i2\pi S Q$. 

In addition, we have terms of the form $\dot{p}{q}$ when hard 
fluctuations are present.
Here, in the spirit of a low-energy theory, we consider the
time derivatives, $\dot{p}_i$'s, to be small. Derivatives of the
hard modes, $\dot{q}_i$'s, will be taken to be doubly small. With these 
assumptions, the leading order single-time-derivative terms are of the 
form $\dot{p}_i q_l$. The coefficients $B_{il}$ can be worked out for 
specific cases, as discussed below.

The combined action for the cluster is given by the sum of Eqs.~\eqref{eq.ENLSM} and \eqref{eq.BerryNLSM}. We may integrate out $q_l$'s, the hard 
fluctuations, to obtain
\begin{eqnarray} S ~=~ i 2\pi S Q ~+~ \int_0^\beta d\tau ~\sum_{i,j=1}^d 
C_{ij} \dot{p}_i \dot{p}_j. \label{eq.actionNLSM} \end{eqnarray}
This can be interpreted as the path integral action of a particle moving on the CGSS, parametrized by $p$'s. The quadratic term, $C_{ij}\dot{p}_i \dot{p}_j$ represents kinetic energy on the CGSS. The coefficients $C_{ij}$ can be determined in terms of $A_{lm}$ and $B_{il}$.
A quantized Berry phase emerges when $QS$ takes half-integer values ($1/2,3/2,5/2,\ldots$). This can be interpreted as $\pi$-flux tubes that are threaded through the space (see examples below), imbuing the particle with Aharonov-Bohm phases.

\section{$XY$ Dimer}
\label{sec.dimer}

\begin{figure}
\includegraphics[width=\columnwidth]{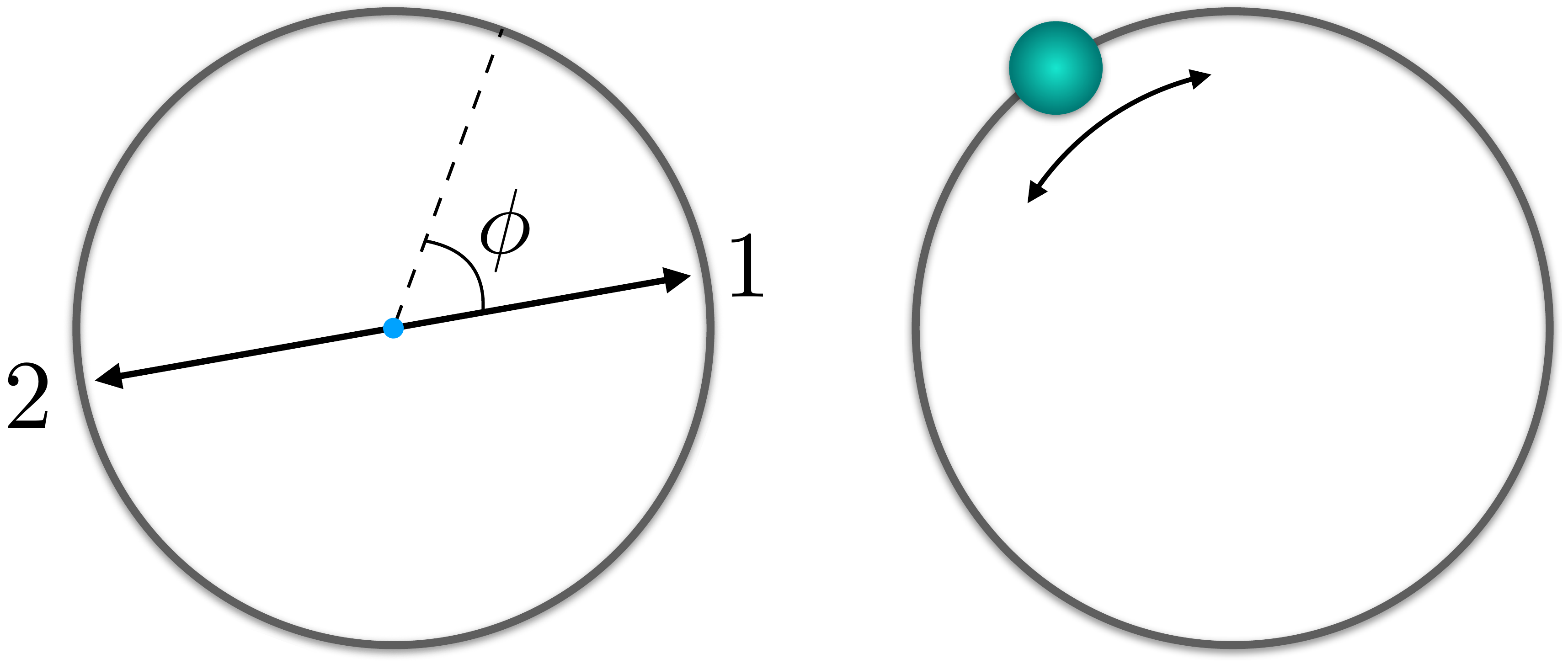}
\caption{Left: Classical ground state of a dimer with both spins lying in the 
$x-y$ plane and pointing in opposite directions. Each ground state is 
parametrized by one angle, $\phi$. Right: A particle moving on a circle.} 
\label{fig.dimergs} \end{figure}

The simplest cluster in our family consists of two spins coupled by an $XY$ bond, with no frustration. The spins are quantum objects with spin quantum number $S$. In the classical limit, in order to minimize energy, the two spins must lie in the $x-y$ plane and point in opposite directions. This ground state is depicted in Fig.~\ref{fig.dimergs}. Any such state can be specified by one angle, $\phi$, representing the position of the first spin. The set of all ground states forms a circle, $\phi \in [0,2\pi)$, with $\phi \equiv \phi \pm 2\pi$. Below, we show that this system maps to a particle moving on a circle as shown in Fig.~\ref{fig.dimergs} (right).

\subsection{Low-energy semiclassical description}
\label{ssec.dimer_nlsm1}

We parametrize the ground states as $\vS_1 = S\hat{n}(\phi)$ and $\vS_2 = -S\hat{n}(\phi)$, where $\hat{n}$ represents a unit vector in the $x-y$ plane. 
The angle $\phi$ represents a dynamical variable that can vary with time.
To describe the low-energy physics, we introduce small fluctuations, in line with Eq.~\eqref{eq.spinnlsm}, 
\begin{eqnarray}\label{eq.fluc_para}
\vec{S}_{1,2} &=& S \frac{ \hat{n}_{1,2} + \vec{r}_{1,2}/S}{\sqrt{1+ 
\vec{r}_{1,2}\cdot \vec{r}_{1,2}/S^2 }}, \end{eqnarray}
with $n_{1,2} = \pm\hat{n}(\phi)$ and $\vec{r}_{1,2} = \vec{l}\pm m \,\hat{z}$. Here, $\vec{l}$ is a three-dimensional vector, representing the magnetization 
of the dimer. It is constrained to be perpendicular to $\hat{n}$, i.e., $\hat{n}\cdot\vec{l} = 0$. 
In addition, we have a staggered moment in the $\hat{z}$ direction, given by $m\hat{z}$. 
Both $\vec{l}$ and $m$ represent hard modes.


\noi \textbf{Berry phase}:
The expression in Eq.~\eqref{eq.BerryNLSM} takes the form
\begin{eqnarray}\label{eq.dimer_Bphase}
2\pi iS Q + iS\int_0^\beta d\tau \left(2\vec{l}\cdot (\partial_\tau
\hat{n}\times \hat{n})\right) = 2i\int_0^\beta d\tau l_z \dot{\phi}.
\end{eqnarray}
Here, the integer $Q$ takes even integer values for any $S$. As the resulting phase is a multiple of $2\pi$, it can be discarded.


\noi \textbf{Energy}:
Using Eq.~\eqref{eq.fluc_para}, the $\mathcal{O}(S^0)$ term in the energy is 
\begin{equation}\label{eq.dimer_energy}
E ~=~ J ~(l_z^2 + m^2 + 2l_x^2 + 2l_y^2). \end{equation} 
Combining the Berry phase and energy terms, after integrating out the
hard modes, the action takes the form
\begin{equation} S^D_{eff} ~=~ \frac{1}{J} \int_0^\beta d\tau ~\dot{\phi}^2.
\end{equation}
This is a well-known form, describing a particle moving on a ring. Here, we interpret $J\equiv 2/(\mu a^2)$, where $\mu$ is the mass of the particle and $a$ is 
the radius of the ring (we will set $a=1$).

\subsection{Comparison with full quantum description}
\label{ssec.dimer_qm}

\begin{figure}
\includegraphics[width=\columnwidth]{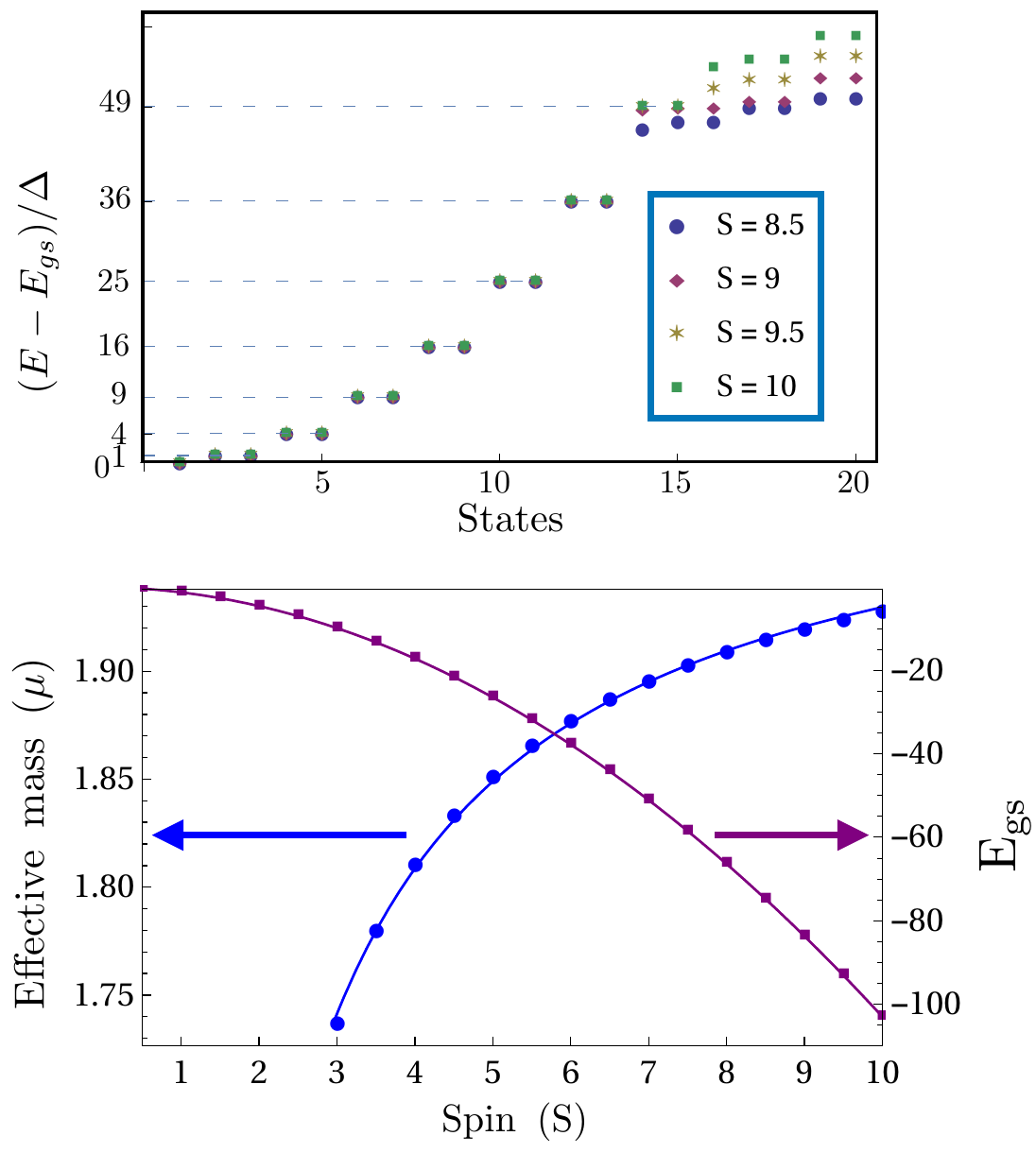}
\caption{Top: Energy spectrum of the $XY$ spin-$S$ dimer obtained numerically, for various $S$ values. The energies have been shifted by $E_{gs}$ (ground state energy) and scaled by $\De$ (energy gap to the first excited state). Energy levels expected for a particle on a ring are marked on the $\hat y$-axis and shown by dashed lines. Bottom: Effective mass $\mu$ (in units of $1/J$) is
plotted as a function of $S$. The data are fit by the plotted curve 
given by $\mu(S)=2.0103-0.80653/S$. Ground state energy $E_{gs}$ (in units of 
$J$) is plotted as a function of $S$ (squares). The data is fit by the curve 
$E_{gs}(S)= - 1.00049 \,S^2 - 0.285\, S - 0.1177$.} \label{fig.dimerspectrum} 
\end{figure}

In order to quantitatively demonstrate the mapping to a particle on a ring, we compare the spectra obtained in the two cases. For a particle on a ring, 
the eigenstates are labeled by angular momentum, $n=0,\pm 1, \pm 2, \cdots $, with the energy being given by $n^2/(2\mu a^2)$. For the spin problem, 
we numerically diagonalize the Hamiltonian to obtain the spectrum. For a spin-$S$ dimer, the Hilbert space dimension is $(2S+1)^2$.
The spectrum, for various $S$ values, is shown in Fig.~\ref{fig.dimerspectrum}. We find excellent agreement with the particle picture. The low-lying energies scale as $n^2$ with ($n=0,\pm 1, \pm 2, \cdots $), with a non-degenerate ground state and doubly degenerate excited states. For example, with $S=10$, we find agreement with this form for the lowest 8 levels (15 states after accounting for degeneracies).

From the numerical data, we extract two quantities for each value of $S$. 

\noi (i) $E_{gs}$, the ground state energy (the lowest eigenvalue of the Hamiltonian): In the $S\rightarrow \infty$ limit, we expect this quantity to give the classical ground state energy, $E_{cl}=-JS^2$. In Fig.~\ref{fig.dimerspectrum} (bottom), we plot the numerically obtained values of $E_{gs}$ vs $S$. The plot shows a fit to a functional form, $E_{gs}=E_2 S^2 + E_1 S + E_0$. The leading term is $E_{gs}\simeq -J S^2$, as expected in the classical limit. We find a significant semiclassical correction in the form of an $\mathcal{O}(S)$ term. We can
quantitatively account for this correction using an analysis
based on the Holstein-Primakoff (HP)
transformation~\cite{Holstein1940,Anderson1952,Kubo1952} (see Appendix~\ref{App.HPdimer}). The HP calculation predicts the $\mathcal{O}(S)$ correction to the ground state energy to be $-(1-\frac{1}{\sqrt{2}})S \simeq -0.2928 S$, which is remarkably close to $- 0.285\, S$, the value obtained from the fitting function $E_{gs}(S)$, given in the caption of Fig.~\ref{fig.dimerspectrum}.

\noi (ii) The scaling factor, $\De$, which is the gap to the first excited state: 
For a particle on a ring, the spacing between energy levels is $(n_2^2 - n_1^2) /(2\mu a^2)$. The scale is the inverse of $2\mu a^2$, twice the moment of inertia of the particle. We extract this quantity from the data in the form of $\De$, the gap to the first excited state. From the preceding path integral derivation, we see that the magnetic coupling $J$ can be interpreted as $2/(\mu 
a^2)$. This equivalence i.e., $1/(2\mu a^2)\simeq J/4, $ is also seen in the HP analysis presented in Appendix~\ref{App.HPdimer}, which predicts a low-lying spectrum given by $Jm^2/4$, where $m=0,\pm 1,\pm 2,\cdots$. 
However, this is a leading order result that agrees with the numerics at large $S$. We find that $\De$ depends on $S$, indicating that the moment of inertia renormalizes with decreasing $S$ (see the fitting function for the effective mass $\mu(S)$ in the caption of Fig.~\ref{fig.dimerspectrum}). 

\section{$XY$ Trimer}
\label{sec.trimer}

\begin{figure}
\includegraphics[width=\columnwidth]{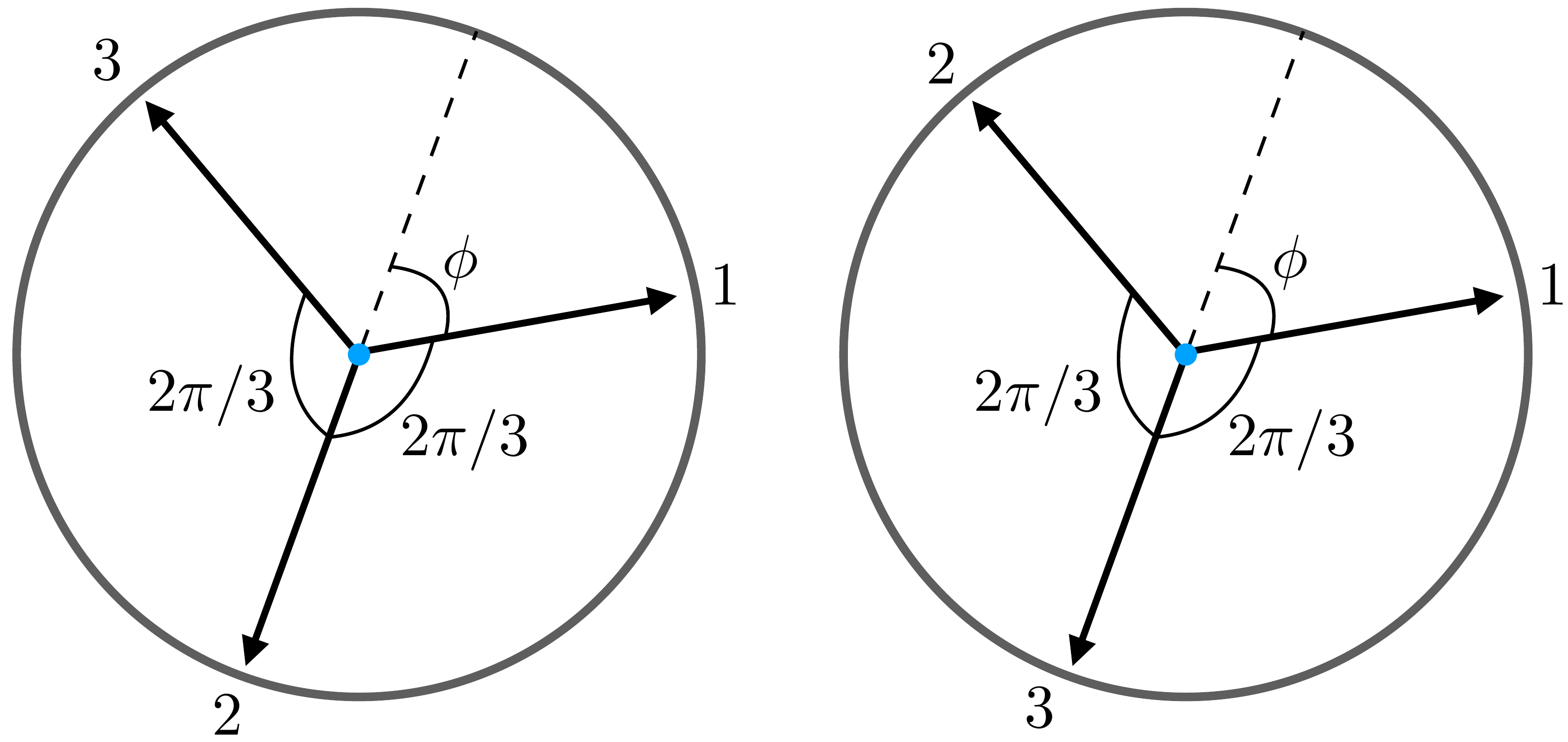}
\caption{Classical ground states of the trimer.} \label{fig.trimergs} 
\end{figure}

With three spins, it is not possible to have every pair of spins anti-aligned. The lowest energy state is obtained by restricting all spins to lie in
the $x-y$ plane, with each pairs of spins subtending an angle of $\pm$120$^\circ$. This can be achieved in the two ways shown in Fig.~\ref{fig.trimergs} -- with spins arranged as $(1,2,3)$ or $(1,3,2)$ in the clockwise direction. In each case, we may perform global spin rotations, captured by the parameter $\phi$ in the figure. Global rotations preserve the handedness of the configuration, i.e., they do not change $(1,2,3)$ to $(1,3,2)$ or vice versa. Thus, the set of all ground states is equivalent to two disjoint circles with a $\mathbb{Z}_2$ parameter labeling the circles. An independent parameter, $\phi$, parametrizes points within each circle. 

\subsection{Low-energy semiclassical description}
\label{ssec.dimer_nlsm2}

We parametrize the ground states as
$\mathcal{S}_{1,\nu} = \hat{n}$,
$\mathcal{S}_{2,\nu} = T_\nu R_z^{\frac{2\pi}{3}}\hat{n}$, and 
$\mathcal{S}_{3,\nu} = T_\nu^{'}R_z^{\frac{4\pi}{3}}\hat{n}$. 
Here, $\nu=1,2$ is the $\mathbb{Z}_2$ order parameter that specifies the 
circle, and $\hat{n}$ is a unit vector in the $x-y$ plane. We have introduced two rotation operators about the $\hat{z}$-axis, $R_z^{\frac{2\pi}{3}}$ and $R_z^{\frac{4\pi}{3}}$, by angles $2\pi/3$ and $4\pi/3$ respectively. The ground 
states on the two circles are distinguished by the
operators $T_1 = I, T_2 = R_z^{\frac{2\pi}{3}}$ and $T_1^{'} = I , T_2^{'} = 
R_z^{\frac{4\pi}{3}}$. We parametrize the spins as
\begin{eqnarray} \vec{S}_{1,2,3}^\nu &=& S ~\frac{ \hat{n}_{1,2,3}^\nu + 
\vec{r}_{1,2,3}^{\phantom{a}\nu} /S}{\sqrt{1+ \vec{r}_{1,2,3}^{\phantom{a}\nu}
\cdot \vec{r}_{1,2,3}^{\phantom{a}\nu} /S^2 }}, \end{eqnarray}
where $\hat{n}_1^\nu = \hat{n}$, $\hat{n}_2^\nu = T^\nu R_z^{\frac{2\pi}{3}} 
\hat{n}$, and $\hat{n}_3^\nu = T^{\nu '}R_z^{\frac{4\pi}{3}}\hat{n}$. 
Hard modes are introduced via the vectors
$\vec{r}_1^{\phantom{a}\nu} = \vec{l}+ m_1 \hat{z}$, $\vec{r}_2^{\phantom{a}\nu} = M_1^\nu\vec{l}+ m_2 \hat{z}$, 
and $\vec{r}_3^{\phantom{a}\nu} = M_2^\nu\vec{l} - (m_1 + m_2)\hat{z}$.
As with the dimer problem, the net magnetization of the trimer is captured by $\vec{l}$. To preserve normalization, we have introduced tensors 
$(M_1^{\nu})^{\al\beta} = \de^{\al\beta} - (T_\nu R_z^{\frac{2\pi}{3}}\hat{n})^{\al}(T_\nu R_z^{\frac{2\pi}{3}}\hat{n}) ^{\beta}$ and $(M_2^{\nu})^{\al\beta} = \de^{\al\beta} - (T_\nu^{'}R_z^{\frac{4\pi}{3}}\hat{n})^{\al}(T_\nu^{'}R_z^{\frac{4\pi}{3}}\hat{n}) ^{\beta} $. These project the $\vec{l}$ vector in each spin onto the plane perpendicular to the ground state vector, in order to satisfy the spin length constraint.

The Berry phase term for the trimer, for each value of $\nu$ in the 
parametrization, comes out to be 
\begin{equation}\label{eq.trimer_Bphase}
\mathcal{S_B} ~=~ 6i\pi S Q ~+~ 3i\int_0^\beta d\tau ~l_z\cdot\dot{\phi}. 
\end{equation}
As the Berry phase only contains the $l_z$ hard mode, we look for $l_z$ terms 
in the energy. Other hard modes do not contribute in the effective action.
For each choice of $\nu$, the energy of the trimer is $E \sim (3J/2) 
l_z^2$. Thus, after integrating out $l_z$, we find the effective action for each $\nu$,
\begin{equation} S^T_{eff} ~=~ 6\pi iS Q ~+~ \frac{3}{2J}\int_0^\beta d\tau ~
\dot{\phi}^2. \end{equation}
This is readily identified as the action of a particle on two disjoint rings, 
due to the two possible values of $\nu$. 

The quantized term in the Berry phase can play a significant role here. To form a closed loop in the ground state space, the three spins must rotate around 
the equator (about the $\hat{z}$-axis) an integer $Q$ number of times. This 
corresponds to sweeping out an area equal to $6\pi S Q$, with $Q\in \mathbb{Z}$. For integer values of $S$, this phase is always a multiple of $2\pi$ that can be discarded. However, for half-integer values of $S$, it gives an odd multiple of $\pi$ when $Q$ is odd. This phase can be adapted to the particle picture as a $\pi$ flux that pierces each ring. When the particle goes around a ring an odd number of times, it picks up an Aharonov-Bohm phase of $\pi$. 

\begin{figure}
\includegraphics[width=\columnwidth]{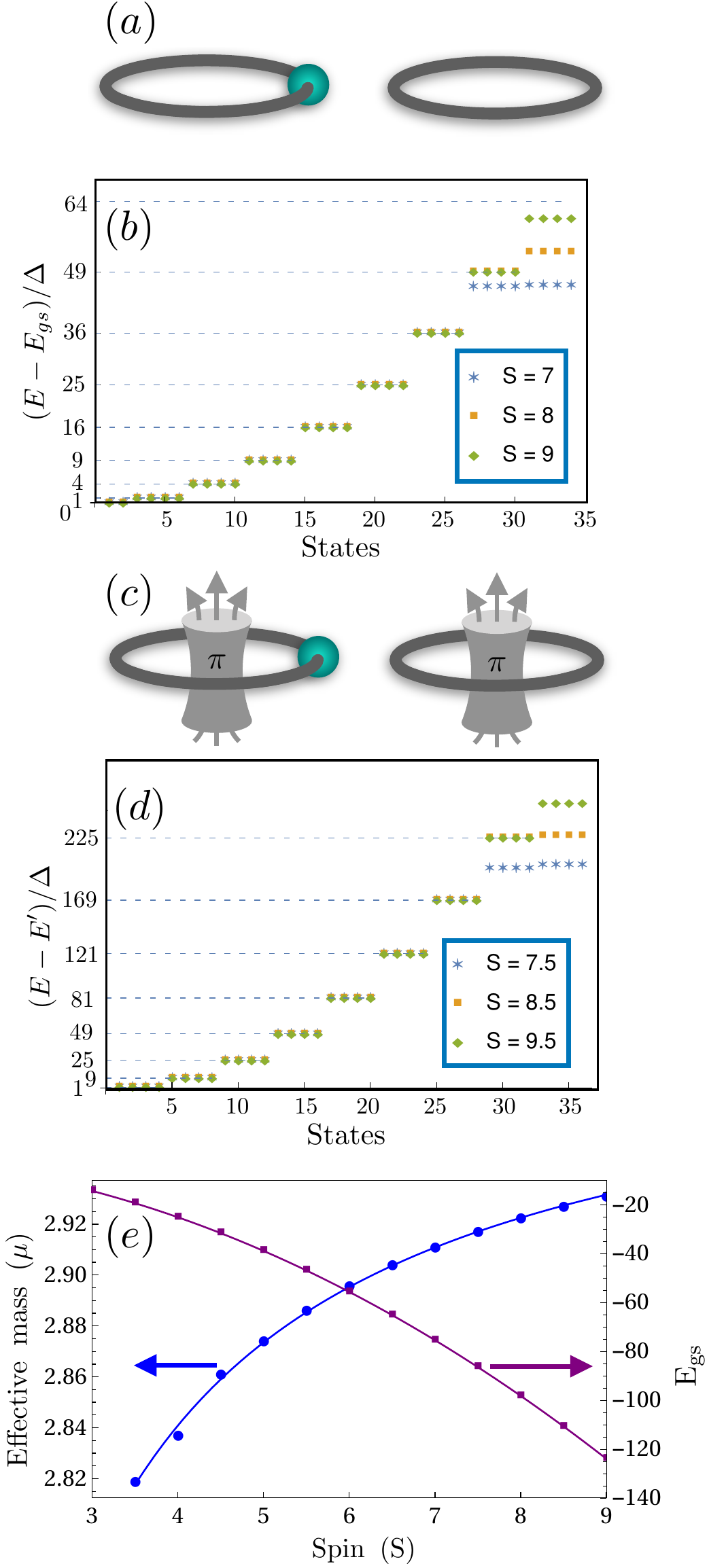}
\caption{(a) Particle on two disjoint rings. (b) The spectrum of the trimer obtained numerically for different integer $S$ values. Energies have been shifted by the ground state energy $(E_{gs})$ and scaled by $\De$ (energy gap to the first excited state). Energy levels for a particle on two disjoint rings are shown by the dashed lines from the $\hat y$-axis. (c) Particle on two disjoint rings with one $\pi$ flux threaded through each ring. (d) Numerically obtained spectrum for the trimer for several half-integer $S$ values. The spectrum has been shifted by $E'$ and scaled by $\De$. $E'$ is chosen to fix the
ground state energy at unity, while $\De$ is chosen to fix the gap to the first excited state at $8$. 
(e) Effective mass $\mu$ (in units of $1/J$)
for various (integer and half-integer) $S$ values shown by circles. 
The data are fit using the $\mu (S)=3.00414 -0.653104/S$. Ground state 
energy $(E_{gs})$ (in units of $J$) for each $S$ is shown using squares. The corresponding fitting function is, $E_{gs}(S)=-1.5002\, 
S^2-0.2701\, S-0.0583$.} \label{fig.trimerspectrum} \end{figure}

\subsection{Comparison with full quantum description}
\label{ssec.trimer_qm}

To quantitatively test the mapping to a particle on two rings, we numerically study the trimer spectrum as a function of $S$. The Hilbert space dimension is $(2S+1)^3$. For integer spins, the mapping is to a particle on two disjoint rings. This has eigenstates labeled by a $\mathbb{Z}_2$ variable and $n$, the angular momentum quantum number. The energy levels are $n^2 /(2\mu a^2)$, with $n=0,\pm 1,\pm2,\cdots$. 
Due to the presence of two disjoint rings, the ground state is doubly degenerate while all excited states are four-fold degenerate. As shown in Fig.~\ref{fig.trimerspectrum}, the numerically obtained energies 
show excellent agreement with this picture.

For half-integer spins, the mapping is to a particle on two rings threaded with $\pi$-fluxes. This has eigenenergies
$(n-1/2)^2 /(2\mu a^2)$, with $n\in \mathbb{Z}$ being the angular momentum. In addition, we have a $\mathbb{Z}_2$ quantum number that picks one of two circles. All low-lying states (including the ground state) are four-fold degenerate. Figure~\ref{fig.trimerspectrum} shows the numerically obtained energies which 
agree well with this picture.

As with the dimer, we extract two quantities from the data, as a function of 
$S$. 

\noi (i) $E_{gs}$, the ground state energy (the lowest eigenvalue of the Hamiltonian): The data is described by a fit of the form $E_{gs} = E_2 S^2 + E_1 S + E_0$. The leading order term is $E_{gs}\simeq -(3/2) J S^2$, consistent with the classical energy of three spins in a $120^\circ$ state. The fit reveals a non-negligible subleading $\mathcal{O}(S)$ correction, emerging from quantum fluctuations. We provide a quantitative explanation for this correction using a 
HP analysis (see Appendix~\ref{App.HPtrimer}). This gives the $\mathcal{O}(S)$ correction to the ground state energy to be $-(1.5 - \sqrt{1.5})S \simeq -0.275\, S$, close to $-0.2701 \,S$, the $\mathcal{O}(S)$ correction from the fitting function for $E_{gs}$ given in the caption of 
Fig.~\ref{fig.trimerspectrum}.

\noi (ii) $\De$: This is taken to be the gap to the first excited level for integer spins, and one-eighth of the gap for half-integer spins.
The path integral derivation above gives the leading order contribution, $\De \sim J/6$ for integer $S$ and $\De \sim J/24$ for half-integer $S$. In the particle picture, this is inversely related to the moment of inertia. We extract this from the $\De$ values. The numerical data shows strong $S$ dependence, indicating that the effective mass of the particle (or more precisely, the moment of inertia) is renormalized by quantum fluctuations for finite $S$ values. The $S$ dependence can be read off from the fitting function, $\mu (S)$, given in the caption of 
Fig.~\ref{fig.trimerspectrum}. The leading order value, $\mu \simeq 3$ 
(i.e., $1/(2\mu a^2) \simeq 1/6$), agrees well with the HP analysis given in 
Appendix~\ref{App.HPtrimer}. The HP low-energy spectrum is given by $Jm^2/6$, where $m=0,\pm 1, \pm 2, \cdots$ for integer $S$ and 
$m=\pm 1/2, \pm 3/2, \cdots$ for half-integer $S$.

\section{Asymmetric $XY$ quadrumer}
\label{sec.asym_quad}

\begin{figure}
\includegraphics[width=\columnwidth]{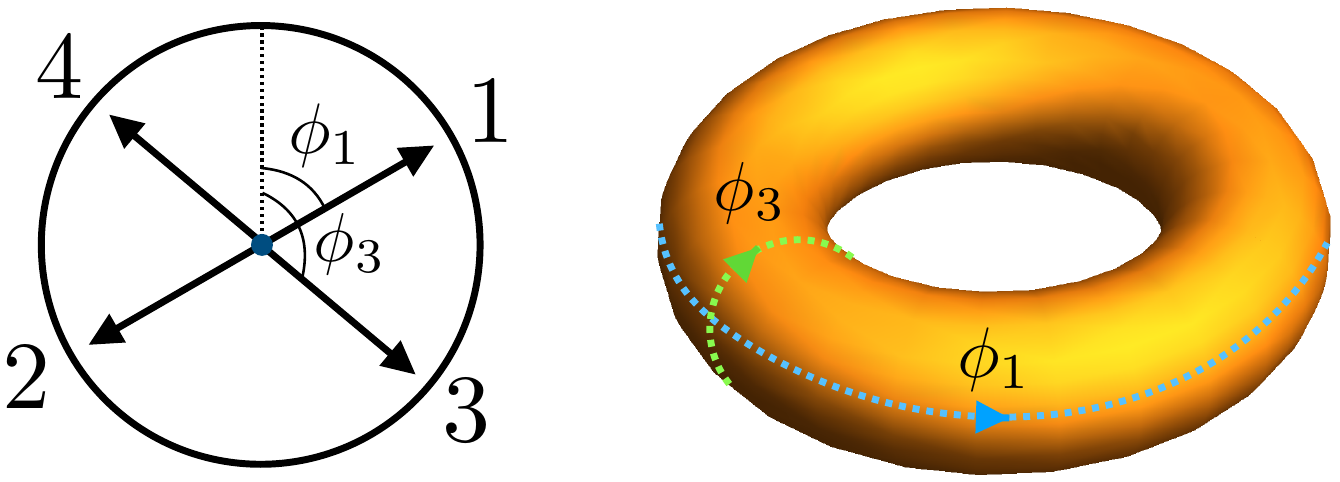}
\caption{Left: Classical ground states of the asymmetric quadrumer 
parametrized by two angles, $\phi_1$ and $\phi_{3}$. Right: The space of 
all ground states, forming a torus.} \label{fig.aquadgs} \end{figure}

\begin{figure}
\includegraphics[width=\columnwidth]{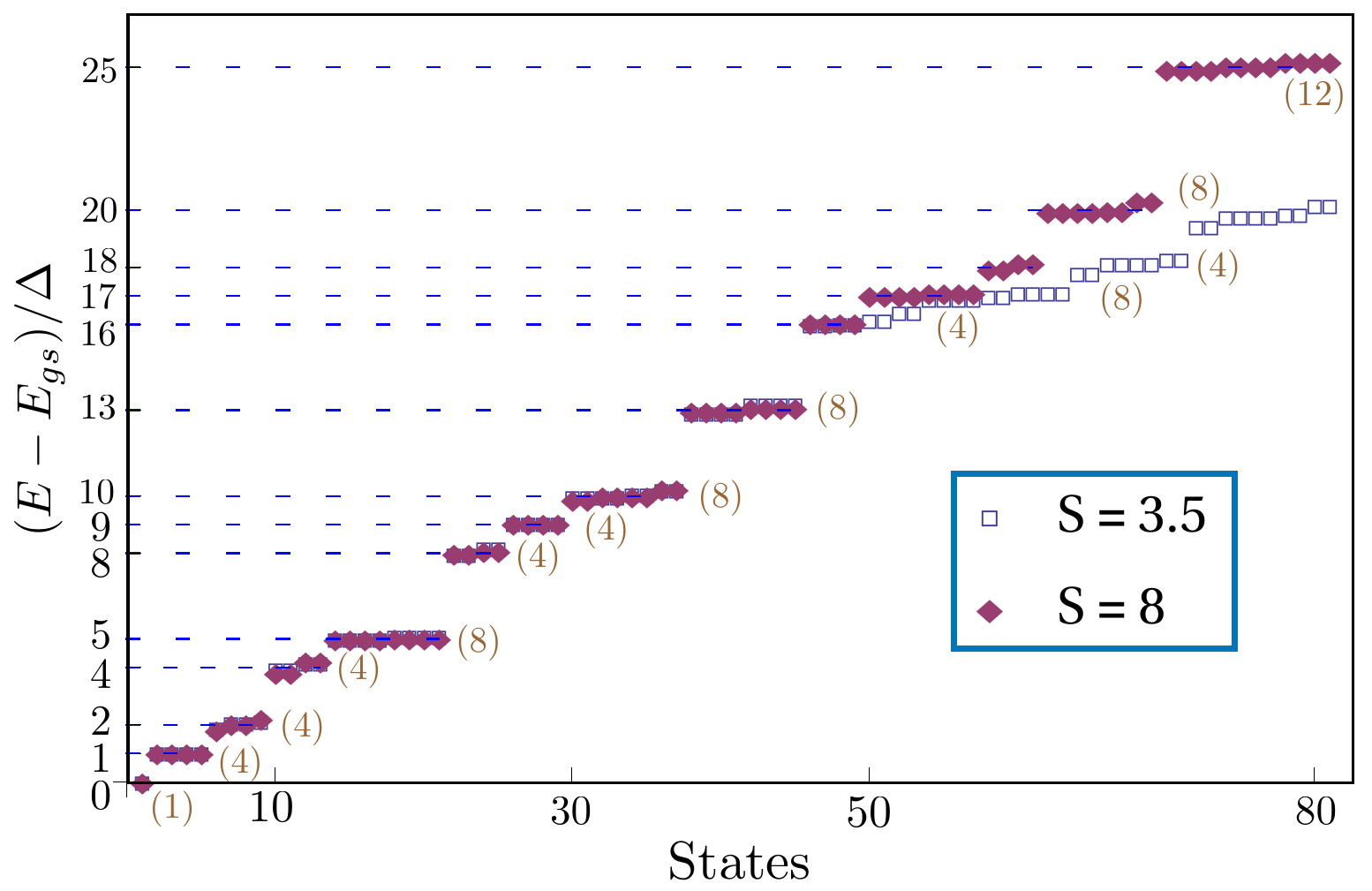}
\caption{Energy spectrum of the $XY$ asymmetric quadrumer obtained numerically for different $S$ values with $\lambda = 2$. Energies are given a shift by the ground state energy $(E_{gs})$ and scaled by $\De$ (so as to fix the gap to the second excited level to 2). The energies of a particle on a torus are marked on the $\hat y$-axis and shown by the dashed lines. The degeneracy of each level is given in 
parentheses.} \label{fig.Aquad_spectrum} \end{figure}

With four spins on a distorted tetrahedron, we have two pairs that have a 
stronger coupling compared to the others. The classical ground state is obtained by anti-aligning these pairs independently.
To see this, we consider the Hamiltonian given by
\begin{eqnarray} 
\nonumber H &=& J \sum_{i<j} ~\vec{S}_i \cdot \vec{S}_j + \lambda J\Big[ \vec{S}_1\cdot\vec{S}_2 + \vec{S}_3\cdot\vec{S}_4 
\Big] \\
&=& \frac{J}{2}
\Big[ \Big(\sum_{i=1}^4 \vec{S}_i \Big)_\parallel^2 - \sum_{i=1}^4 (\vec{S}_{i})_\parallel^2\Big]
+ \lambda J \Big[ \vec{S}_1\cdot\vec{S}_2 + \vec{S}_3\cdot\vec{S}_4 
\Big].\phantom{abcd}
\label{eq.asym_Hmlt}
\end{eqnarray}
Here, $\vec{A}\cdot \vec{B} \equiv A_x B_x + A_y B_y$ denotes an $XY$ dot product and $(\vec{A})_\parallel^2 \equiv A_xA_x + A_yA_y$. 
The first term is minimized when the in-plane components of the spins add to zero, while the second term forces all spins to lie in the $x-y$ plane. Taken together with the $\lambda$ term, we deduce that classical ground states are as shown in Fig.~\ref{fig.aquadgs} (left). Pairs of spins, $(1,2)$ and $(3,4)$, are anti-aligned. The relative angle between the two pairs (e.g., between $\vec{S}_1$ and $\vec{S}_3$) is not constrained. 
The set of all such states can be described by two angles, $\phi_1$ and $\phi_3$. The former denotes the position of the first spin. This immediately specifies the second spin, which is anti-aligned with respect to the first. The latter fixes the third spin, and thereby the fourth as well. These two parameters are angle variables, periodic with domain $[0,2\pi)$. 

The CGSS is equivalent to a torus as shown in Fig.~\ref{fig.aquadgs} (right). This is a two-dimensional manifold, with greater complexity than the dimer and trimer discussed above. This represents a qualitative change as the dimer and trimer have ground states that are related by global rotations about the $\hat{z}$-axis, a symmetry of the Hamiltonian. For the asymmetric quadrumer, the CGSS (2D) is bigger than the space of symmetries (1D). This represents an 
`accidental degeneracy' that is not protected by symmetry, a classic feature of frustrated magnets. As a consequence, we have the possibility of order by disorder. We discuss this using a HP approach below. As order by disorder effects are negligible for sufficiently large $S$, we first describe an effective theory considering the full CGSS. We present numerical data which are found to be in agreement with our analysis.

\subsection{Low-energy semiclassical description}
\label{ssec.asym_quad_nlsm}

The classical ground states are given by 
\begin{eqnarray} \vS_1 &=& \hat{n}_1(\phi_1), ~~~~\vS_2 = -\hat{n}_1(\phi_1),
\non \\
\vS_3 &=& \hat{n}_2(\phi_3), ~~~~\vS_4 = -\hat{n}_2(\phi_3). \end{eqnarray}
Here $(\vS_1, \vS_2)$ and $(\vS_3, \vS_4)$ form two separate rods, composed of oppositely aligned spins. We parametrize 
the hard fluctuations as follows, 
\begin{eqnarray}\label{eq.asym_quad_hard_fluc}
\vS_1 &=& \frac{\hat{n}_1 + \frac{\vec{l}_1 + m_1\hat{z}}{S}}{\sqrt{1+\frac{(\vec{l}_1 + m_1\hat{z})^2}{S^2}}},\phantom{a}
\vS_2 = \frac{-\hat{n}_1 + \frac{\vec{l}_1 - m_1\hat{z}}{S}}{\sqrt{1+\frac{(\vec{l}_1 - m_1\hat{z})^2}{S^2}}},\non \\
\vS_3 &=& \frac{\hat{n}_2 +\frac{\vec{l}_2 + m_2\hat{z}}{S}}{\sqrt{1+\frac{(\vec{l}_2 + m_2\hat{z})^2}{S^2}}},\phantom{a}
\vS_4 = \frac{-\hat{n}_2 + \frac{\vec{l}_2 - m_2\hat{z}}{S}}{\sqrt{1+\frac{(\vec{l}_2 - m_2\hat{z})^2}{S^2}}}. \end{eqnarray}
The magnetization of the first rod is $\vec{l}_1$ and that of the second is $\vec{l}_2$. The staggered $z$-magnetization of the rods is denoted by $m_{1/2}$. 
\\
\textbf{Berry Phase:}
Using the parametrization given in Eq.~\eqref{eq.asym_quad_hard_fluc}, the Berry phase is found to be
\begin{eqnarray}
\mathcal{S}_B &=& 4\pi iSQ_1 ~+~ 4\pi iSQ_2 ~+~ 2i \int_0^\beta d\tau ~
(l_{1z}\dot{\phi}_1 + l_{2z}\dot{\phi}_3) \non \\
&=& 2i ~\int_0^\beta d\tau ~( l_{1z}\dot{\phi}_1 + l_{2z}\dot{\phi}_3). 
\end{eqnarray} 
Here, $Q_1, Q_2$ are integers. The quantized Berry phase is a multiple of $2\pi$ for any $S$, and can be discarded.\\
\textbf{Energy:}
As in the trimer, we look for terms which include $l_{1z}$ or $l_{2z}$. We 
find that the energy is $E =(1+\lm)J\int_0^\beta d\tau (l_{1z}^2 + l_{2z}^2)$. After integrating out $l_{1z}$ and $l_{2z}$, the effective action is found to be
\begin{equation}
\mathcal{S}^{asymQuad}_{eff} ~=~ \frac{1}{(1+\lm)J} ~\int_0^\beta d\tau ~
(\dot{\phi}_1^2 + \dot{\phi}_3^2). \end{equation}
This is precisely the action for a particle moving on a torus. We find 
no distinction between half-integer and integer cases. 

\subsection{Comparison with full quantum description}
\label{ssec.asym_quad_qm}

We obtain the spectrum of the asymmetric quadrumer numerically. The Hilbert space dimension is $(2S+1)^4$, which grows rapidly with $S$. To perform diagonalizations for large values of $S$, we use the
symmetry under spin rotations about the $\hat{z}$-axis. A particle moving on a torus is known to have the spectrum $(n_1^2+ n_2^2) /(2\mu a^2)$, where $a$ is the radius of the torus in both directions. The ground state is non-degenerate, while excited states are typically degenerate. The numerically obtained spectrum shows excellent agreement with this picture. This is shown in Fig.~\ref{fig.Aquad_spectrum}. The $\hat y$-axis labels (marked by dashed lines) are the known energy levels for a particle on a torus. The expected degeneracy of each level is shown in parentheses. The numerical data shows 
excellent agreement. For instance, with $\lm=2$, we find $9$ levels ($45$ states) in agreement for $S=3.5$. The agreement improves for larger spins with $14$ matching levels ($81$ states) for $S=8$.

\subsection{Holstein-Primakoff analysis}
\label{ssec.asym_quad_HP}

As discussed above, the asymmetric quadrumer allows for the possibility of order by disorder. To see this, we undertake a HP
analysis~\cite{Holstein1940,Anderson1952,Kubo1952}. We consider small fluctuations about a classical ground state described by ${\vec S}_j = S (\cos \phi_j, \sin \phi_j, 0)$. In line with Fig.~\ref{fig.aquadgs}, we take $\phi_2 = \phi_1+\pi$ and $\phi_4 = \phi_3+\pi$. We introduce Holstein-Primakoff creation
and annihilation operators, 
\begin{eqnarray}
\cos \phi_j S_j^x + \sin \phi_j S_j^y &=& S - a_j^\da a_j, \non \\
- \sin \phi_j S_j^x + \cos \phi_j S_j^y &\simeq& \sqrt{\frac{S}{2}} (a_j + 
a_j^\da), \non \\
S_j^z &\simeq& - i \sqrt{\frac{S}{2}} ( a_j - a_j^\da).
\label{eq.hp1} \end{eqnarray}
We have ignored $\mathcal{O}(1/S)$ and higher order terms, assuming a 
large value of $S$ as appropriate for the semiclassical limit.
We now introduce dimensionless and canonically conjugate operators 
$x_j$ and $p_j$ (satisfying $[x_j, p_k] = \de_{jk}$), such that
\begin{eqnarray} a_j = \frac{1}{\sqrt{2}} (x_j + i p_j) ~~~{\rm and}~~~
a_j^\da = \frac{1}{\sqrt{2}} (x_j - i p_j). \label{eq.HP_xp} \end{eqnarray}
In this language, Eqs.~\eqref{eq.hp1} take the form
\begin{eqnarray} \cos \phi_j S_j^x + \sin \phi_j S_j^y &=& S - \frac{1}{2} 
(p_j^2 + x_j^2 - 1), \non \\
- \sin \phi_j S_j^x + \cos \phi_j S_j^y &=& \sqrt{S} x_j, \non \\
S_j^z &=& \sqrt{S} p_j. \label{eq.hp3} \end{eqnarray}
Taking the values of the $\phi_j$ angles appropriate for a classical ground state, we write the Hamiltonian in terms of $x_j$'s and $p_j$'s. Keeping terms 
only up to second order in these operators, we obtain an expression which
contains terms up to order $S$. Diagonalizing this Hamiltonian gives the 
ground state energy and the low-energy (HP) spectrum. The HP spectrum 
typically has two parts: free particles and simple harmonic oscillators
(SHO's). The ground state energy is obtained by including the leading quantum correction, namely, the zero point energies of the SHOs. We find 
\begin{eqnarray} H &=& - 2J S^2 (1 + \lm) - 2 J S (1 + \lm) \non \\
&+& \frac{JS}{2} \Big[ (1 + \lm) (p_1^2 + p_2^2 + p_3^2 + p_4^2) \non \\
&+& (1 + \lm) (x_1^2 + x_2^2 + x_3^2 + x_4^2 -2 x_1 x_2 -2 x_3 x_4) \non \\
&+& 2 \cos \phi_{31} (x_1 - x_2) (x_3 - x_4)\Big], \label{ham42} \end{eqnarray}
where $\phi_{31} = \phi_3-\phi_1$. We diagonalize this by defining the
following linear combinations,
\begin{eqnarray} P &=& (p_1 + p_2 + p_3 + p_4)/{2}, \non \\
p_a &=& (p_1 + p_2 - p_3 - p_4)/{2}, \non \\
p_b &=& (p_1 - p_2 + p_3 - p_4)/{2}, \non \\
p_c &=& (p_1 - p_2 - p_3 + p_4)/{2}, \end{eqnarray}
and the corresponding canonically conjugate variables $X, x_a, x_b$ and 
$x_c$. The operator $P$ is related to the total $S^z \equiv \sum_{j=1}^4 S_i^z$,
as $P = S^z/(2\sqrt{S})$. The Hamiltonian then takes the form
\begin{eqnarray} H &=& - 2J S^2 (1 + \lm) - 2 J S (1 + \lm) 
\non \\
&& + JS \Big[ \frac{(1 + \lm)}{2} (P^2 + p_a^2 + p_b^2 + p_c^2) \non \\
&& + (1 + \cos \phi_{31} + \lm) x_b^2 + (1 - \cos \phi_{31} + \lm) x_c^2\Big]\!. \phantom{ab}~\label{ham43} \end{eqnarray}
We thus have two free particles described by $P^2$ and $p_a^2$, and two SHOs
described by $(p_b,x_b)$ and $(p_c,x_c)$. The latter have frequencies 
\begin{eqnarray} \om_{b,c} &=& JS \sqrt{2 (1+\lm) (1 \pm \cos \phi_{31} 
+ \lm)}.\label{ombc} \end{eqnarray}
The ground state energy is given by
\begin{eqnarray} E_0 = - 2J S^2 (1 + \lm) - 2 J S (1 + \lm) + 
\frac{1}{2} (\om_b + \om_c). \label{e41} \end{eqnarray}
The complete energy spectrum is given by
\begin{eqnarray} E = E_0 + n_b \om_b + n_c \om_c + \frac{JS}{2} 
(1 + \lm) (r^2 + s^2), \label{e42} \end{eqnarray}
where $n_b, n_c = 0, 1, 2, \cdots$ are SHO quantum numbers and $r, s$ 
are eigenvalues of
$P$ and $p_a$ respectively. To find the possible values of $r$ and $s$,
we note that $P = S^z/(2 \sqrt{S})$ and $p_a = (S_1^z + S_2^z - S_3^z - S_4^z)/
(2 \sqrt{S})$. From the rules of addition of quantum spins, we see that the 
eigenvalues of $P$ and $p_a$ will be of the form $m/(2\sqrt{S})$
and $n/(2\sqrt{S})$ respectively, where $m,n = 0, \pm 1, \pm 2, \cdots$, regardless of 
whether $S$ is an integer or a half-integer. Hence, the energy spectrum is
\begin{eqnarray} E = E_0 + n_b \om_b + n_c \om_c + \frac{J}{8} 
(1 + \lm) (m^2 + n^2). \label{e43} \end{eqnarray}
The lowest branch of excitations corresponds to $n_b = n_c = 0$, with both SHOs in their ground states. The energy then reduces to the spectrum of a particle moving on a direct product of two circles, i.e., a torus $T^2$. 

This analysis agrees with the effective theory derived above in that it maps to a particle on a torus. 
However, there is a crucial difference. In the HP approach, the ground state energy includes a zero point correction that depends on $\phi_{31}$, a parameter that is not a symmetry variable. This is a manifestation of order by disorder. Notably, the lowest zero point energy is achieved when $\phi_{31}=0$ or $\pi$, representing two distinct collinear ground states. Taking the HP 
results at face value, we would conclude that the system is confined to two collinear sectors. In this case, the variable $p_a$ would not correspond to a true free particle as it moves the system away from collinearity (presumably, a potential energy term in $x_a$ may emerge from higher order terms). We would then
expect the low-energy spectrum to resemble a particle on two disjoint rings (two collinear states corresponding to $\phi_{31}=0$ or $\pi$). This would lead to two-fold degeneracies in each low-lying level. 

However, our numerical results show that this is not the case. For reasonably large $S$ (e.g., for $S=3.5$ as shown in Fig.~\ref{fig.Aquad_spectrum}), the spectrum shows excellent agreement with a particle on a torus. We conclude that the order by disorder potential, being a $1/S$ correction, does not play a role for sufficiently large $S$. We provide further evidence for this in Sec.~\ref{sec.ObSED} below. The irrelevance of the order by disorder potential is intimately tied to the zero-dimensional character of our problem. Order by disorder is usually discussed for magnets in the thermodynamic limit, where the zero point energy receives contributions from a large number of modes. This can amplify the quantum correction and `select' certain ground states. Here, as shown by our numerics, a good description of the low-energy physics is obtained by neglecting this effect.

\section{$XY$ Quadrumer}
\label{sec.quad}

\begin{figure}
\includegraphics[width=\columnwidth]{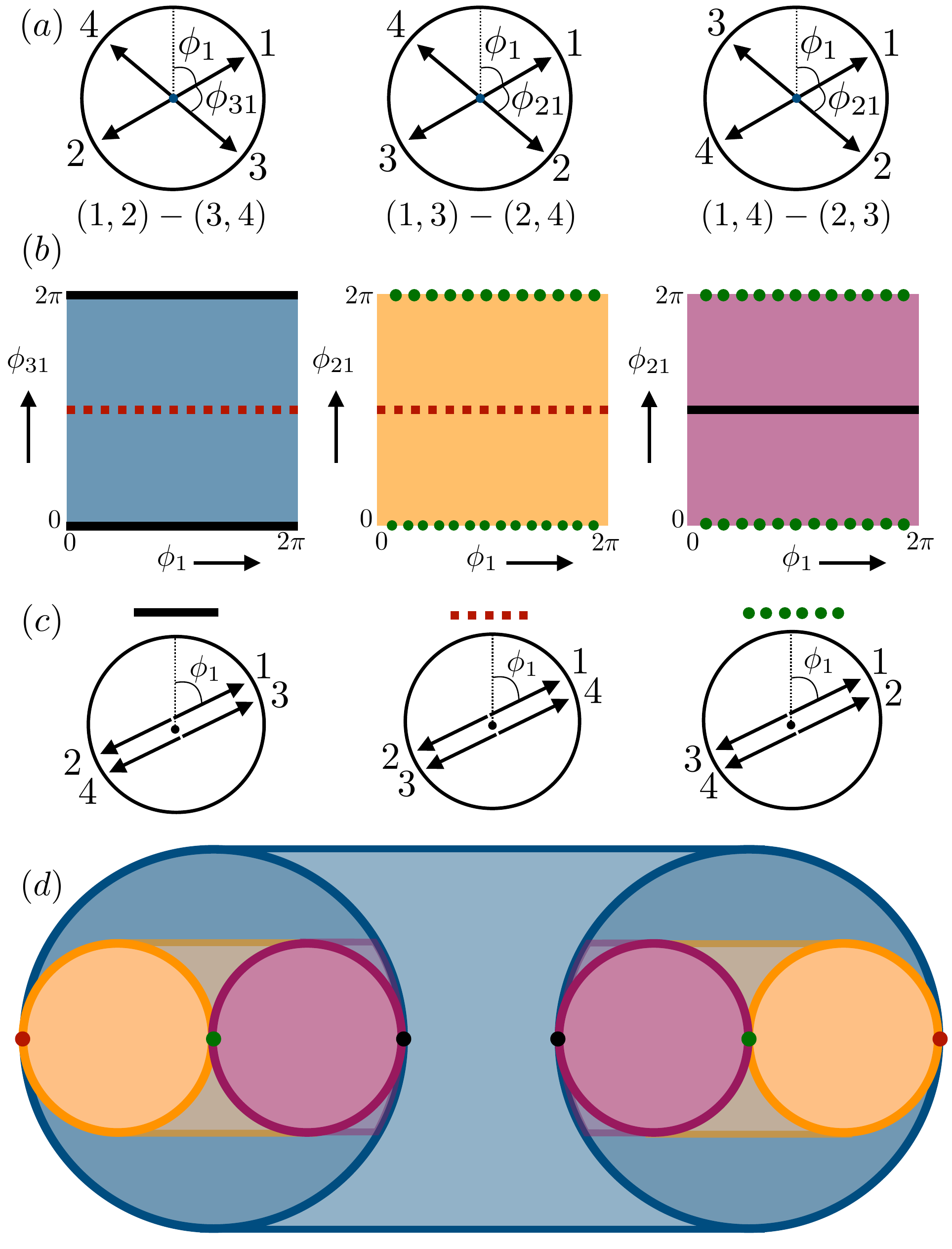}
\caption{Ground state space of symmetric quadrumer. (a) Three possible ways to minimize classical energy, each with two pairs of spins anti-aligned. Each configuration is specified by two angles as shown. 
(b) The space of classical ground states forming three tori. Each torus corresponds to one of the configurations shown above.
(c) Collinear states that appear as lines on the tori, with each line shared by two tori. (d) Cross-section view of the ground state space. We have three tori, with each pair of tori touching along a line.}
\label{fig.quadgs} \end{figure}
The classical ground state space of the (symmetric) quadrumer is qualitatively different from the preceding cases. To minimize the classical energy, we must have all spins lying in the $x-y$ plane, with the vector sum of 
the spins being zero. This is directly seen by setting $\lambda=0$ in Eq.~\eqref{eq.asym_Hmlt}. 
In order for four coplanar vectors to add to zero, we must necessarily have two pairs of anti-aligned spins. This can be seen by placing the four vectors in a head-to-tail arrangement. As the vectors have uniform length and lie on the same plane, their sum can only be zero if they form the sides of a rhombus. The opposite sides of the rhombus correspond to anti-aligned spins.
This leads to three distinct possibilities as shown in Fig.~\ref{fig.quadgs} (a). Consider the first spin, $\vec{S}_1$. It can be anti-aligned with respect to $\vec{S}_2$, $\vec{S}_3$ or $\vec{S}_4$. We denote these three possibilities as $(1,2)-(3,4)$, $(1,3)-(2,4)$ and $(1,4)-(2,3)$. 

States in $(1,2)-(3,4)$ have $\vec{S}_1=-\vec{S}_2$ and $\vec{S}_3=-\vec{S}_4$. To represent a state from this particular family, we need two independent parameters. We first fix $\vec{S}_1$ using an angle $\phi_1$, defined with respect to an arbitrary reference point. Clearly, $\phi_1\in (0,2\pi]$ is an $S^1$ variable, with the periodicity of a circle, i.e., $\phi_1+2n\pi \equiv \phi_1$, where $n$ is an integer.
This immediately fixes $\vec{S}_2$ to be opposite to $\vec{S}_1$. We introduce a second parameter, $\phi_{31}$ to fix the deviation of $\vec{S}_3$ from $\vec{S}_1$. We assume that this angle is measured in the clockwise direction. Once again, $\phi_{31}\in(0,2\pi]$, with the periodicity of $S^1$. Thus, all states in $(1,2)-(3,4)$ can be represented by two parameters, $\phi_1$ and $\phi_{31}$. This space forms a torus, $S^1 \otimes S^1$. In Fig.~\ref{fig.quadgs} (b) (left), we represent this as a square with periodic boundary conditions in 
the horizontal and vertical directions.

Similarly, the space $(1,3)-(2,4)$ is also a torus, parametrized by $\phi_1$ and $\phi_{21}$. Here, $\phi_{21}$ is the angular displacement from $\vec{S}_1$ to $\vec{S}_2$. This is depicted as the central square in Fig.~\ref{fig.quadgs} (b). The third space, $(1,4)-(2,3)$, is likewise a torus parametrized by $\phi_1$ and $\phi_{21}$. It is shown as the square on the right in Fig.~\ref{fig.quadgs} (b).

Naively, the ground state space appears to be three distinct tori. However, there is a subtlety. In each of the three tori, there are two special subsets which contain collinear states. For example, in $(1,2)-(3,4)$, $\phi_{31} = 0,\pi$ correspond to collinear states. These are shown as the
black solid line and the red dotted line in the left square in Fig.~\ref{fig.quadgs} (b). We see that each torus similarly has two special lines. A deeper inspection reveals that the line with $\phi_{31} = 0$ in $(1,2)-(3,4)$ is, in fact, the same as that with $\phi_{21}=\pi$ in $(1,4)-(2,3)$ in Fig.~\ref{fig.quadgs} (b) (right). These are both shown as black solid lines in the figure. Similarly, we note that there are two other pairs of lines that are identical. These lines correspond to three possible collinear states as shown in Fig.~\ref{fig.quadgs} (c). Apart from these lines, there is no state in one torus which also exists in another torus. 

From these arguments, we are able to see the deeper structure of the ground state space. It is composed of three tori, with each pair of tori overlapping along a circle (a line with periodic boundary conditions). This leads to the geometry shown in Fig.~\ref{fig.quadgs} (d) as a cross-section. We have embedded the tori in three dimensions to bring out the connectivity of the space. We see that two tori are enclosed within a third larger torus such that each one touches the larger torus along a circle. The two tori themselves touch along a circle, as shown in the figure.

This ground state space is qualitatively different from the cases discussed in the sections above.
It is a \textit{non-manifold}, as it does not have a
well-defined dimensionality at the common lines where two tori touch. In other words, we cannot define derivatives at the singular lines. This crucial difference precludes a path integral-based low-energy effective theory as laid out in Sec.~\ref{sec.nlsm} and applied to the dimer, trimer and the asymmetric quadrumer.
Nevertheless, we conjecture that the general principle applies here as well, i.e., the low-energy physics of the XY quadrumer maps to that of a particle moving on the non-manifold CGSS. As discussed below, we find strong numerical evidence that this is indeed true. 

 To study a particle in this space, we use a tight-binding approach with a suitable discretization. We discuss the case of integer $S$ and half-integer $S$ separately, due to differences in the
Berry phase structure. 
In Appendix~\ref{App.tangent}, we provide a rigorous discussion of the nature of the CGSS and its tangent spaces at different points. This brings out the non-manifold character of the space and the suitability of the tight-binding model discussed below.

\subsection{Tight-binding approach for integer spins}
\label{ssec.quad_tb_int}

\begin{figure}
\includegraphics[width=\columnwidth]{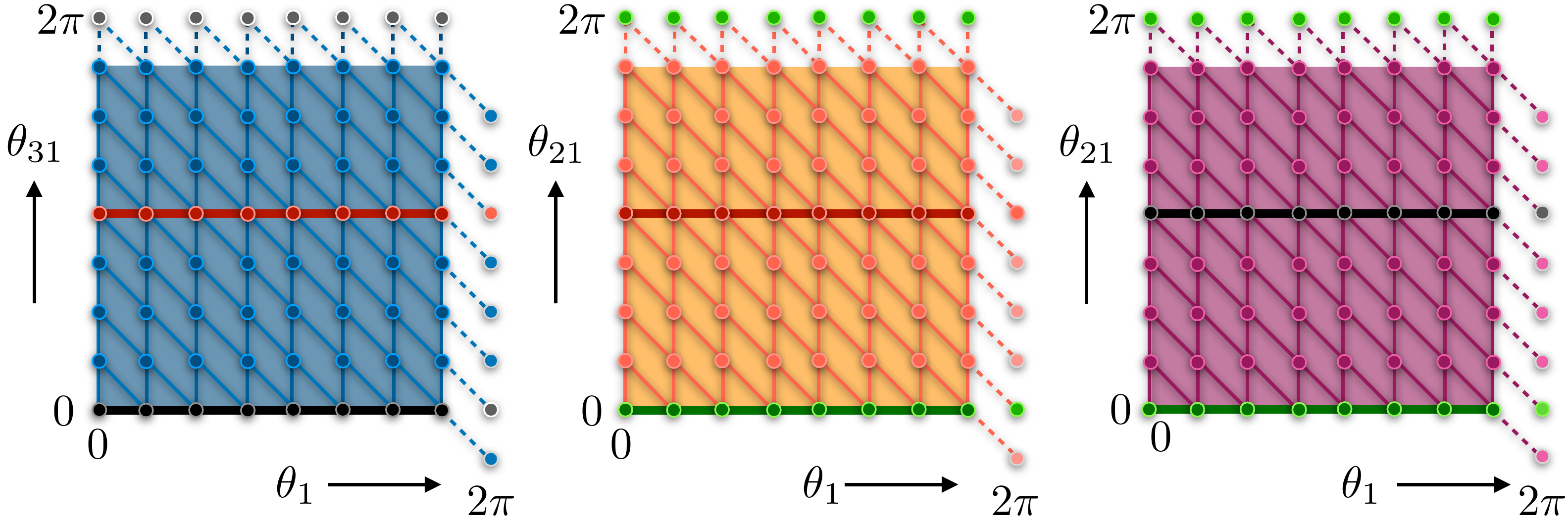}
\caption{Tight-binding mesh that provides a discretization of the quadrumer ground state space. The figure shows an $8 \times 8$ mesh on each torus. The 
dashed lines enforce periodic boundary conditions.} \label{fig.quad_mesh} 
\end{figure}

We discretize the CGSS using the mesh shown in Fig.~\ref{fig.quad_mesh}. This allows for a tight-binding description with the particle hopping from one node to another. We allow hopping along vertical and diagonal bonds with equal amplitudes, with no hopping in the horizontal direction. The bonds connect nodes that are closest to each other in terms of the
displacements of the four spins in the quadrumer (see Appendix~\ref{App.dist_quad_gspace}). We have two free parameters: $L$, the linear size of each torus, and $t$, the hopping strength. In order to capture the connectivity of the space, we identify common lines between tori. For example, in Fig.~\ref{fig.quad_mesh}, 
the central lines of the left and center tori are assumed to have the same physical nodes. A particle on such a node can hop to either torus. With this identification, the number of distinct lattice points is $3L(L-1)$. This sets the size of the Hilbert space for the tight-binding problem. The 
numerically obtained low-energy tight-binding spectrum is shown in Fig.~\ref{fig.tb_intS} for $(t,L)=(0.954419,12)$ and $(2.954755,22)$.

\subsection{Comparison with full quantum description for integer spins}
\label{ssec.quad_qm_int} 

We solve the spin problem for the quantum $XY$ quadrumer using exact diagonalization. The Hilbert space is $(2S+1)^4$-dimensional, intractably large even for intermediate values of $S$. We use two symmetries to reduce the size
of the Hamiltonian: (a) spin rotations about $\hat{z}$, and (b) cyclic permutations, i.e., symmetry under $\vec{S}_1 \rightarrow \vec{S}_2 \rightarrow \vec{S}_3\rightarrow \vec{S}_4\rightarrow \vec{S}_1$. The former divides the Hilbert space into sectors with fixed total $S_z$. The latter reduces it further into angular momentum sectors. These symmetries allow us to work with large spins, up to $S\lesssim 19$.

\begin{figure}
\includegraphics[width=\columnwidth]{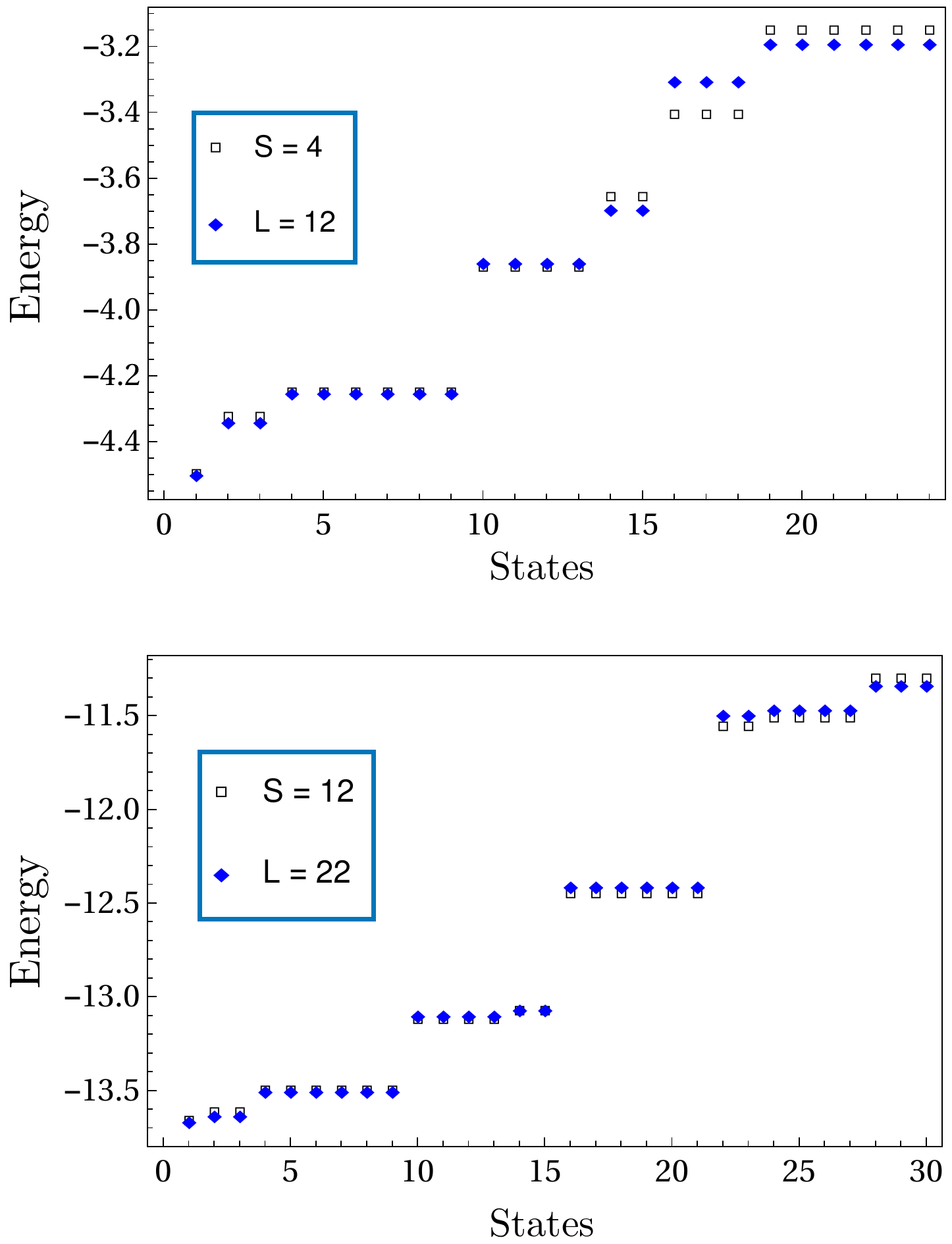}
\caption{Comparison of the spin (empty squares) and tight-binding spectra (solid diamonds) for integer $S$. Spin spectrum has been given a shift such that the ground state energies of spin and tight-binding spectra become the same. }
\label{fig.tb_intS} \end{figure}

Remarkably, the numerically obtained low-energy spectra show excellent agreement with tight-binding results using two fitting parameters: $L$ (torus size) and $t$ (hopping amplitude). This can be seen from Fig.~\ref{fig.tb_intS} which shows the spectra for $S=4$ and $S=12$. We use the following fitting procedure for each $S$. We first fix $L$ (torus size) at an arbitrary value. The hopping $t$ sets the overall energy scale in the tight-binding problem. We fix this scale by fitting the energy gap to the third excited level, comparing the tight-binding value to that from exact diagonalization of the spin Hamiltonian. The choice of the third level provides a large numerical value of the gap, allowing for a robust fit. We now compare the full tight-binding spectrum with that from exact diagonalization. We count the number of low-energy states that match -- we say a state matches if it has the same degeneracy in both approaches, even if the numerical energy values differ.
For instance, for $S=3$ with $L=10$, we find that the lowest 7 levels (24 states after accounting for degeneracies) match. We repeat this procedure for many $L$'s, choosing the value which gives us the highest number of matching states. This procedure gives reasonable values for the fitting parameters as well as good quantitative agreement between spectra. The obtained fit parameters are shown in Table~\ref{tab.intspin}.

We find that the number of matched levels increases with $S$, indicating that the mapping to the tight-binding problem improves when approaching the classical limit. Both $L$ and $t$ increase with $S$. Larger $L$ suggests that more semiclassical orbits are accessed by the particle. At the same time, we find reasonable agreement even for small values of $S$, starting from $S=1$. 

\begin{table}
\begin{centering}
\begin{tabular} { | c | c | c | c |} 
 \hline
Spin & $L$ & Hopping & Levels matched \\ 
$S$ & & $t$ & (No. of states) \\
\hline
\hline
1 & ~~6~~ & ~0.327719~ & 3 ~~(9) \\
2 & 8 & 0.445006 & 4 ~(13) \\ 
3 & 10 & 0.662232 & 7 ~(24) \\
4 & 12 & 0.954419 & 7 ~(24) \\
5 & 12 & 0.891563 & 7 ~(24) \\
6 & 14 & 1.23335 & 8 ~(30) \\
7 & 16 & 1.643835 & 8 ~(30) \\
8 & 18 & 2.110122 & 8 ~(30) \\
9 & 20 & 2.649102 & 8 ~(26) \\
10& 20 & 2.518818 & 8 ~(26) \\
11& 22 & 3.084026 & 9 ~(30) \\
12& 22 & 2.954755 & 9 ~(30) \\
\hline
\end{tabular}
\caption{Comparison of spin and tight-binding spectra for integer $S$. The 
columns show the $L$ and $t$ parameters as obtained by our fitting procedure. The last column shows the number of low-energy levels (and states, after accounting for degeneracies) that match in the two approaches.} \label{tab.intspin}
\end{centering}
\end{table}

\subsection{Tight-binding approach for half-integer spins}
\label{ssec.quad_qm_hint}

\begin{figure}
\includegraphics[width=\columnwidth]{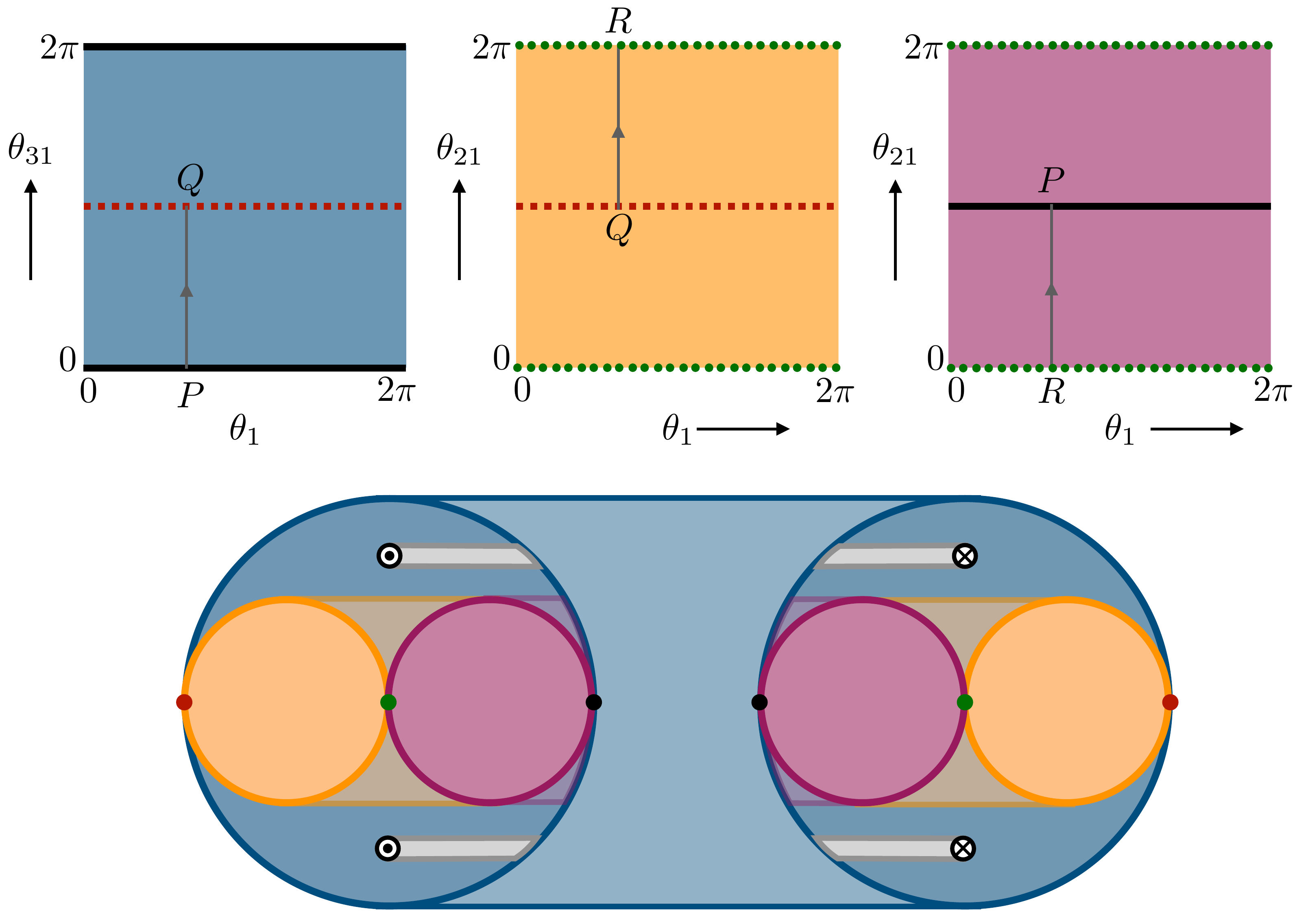}
\caption{Top: Example of a non-trivial path that incurs a $\pi$ Berry phase 
for half-integer values of $S$. Bottom: Cross-section view, with $\pi$-flux 
tubes inserted to account for the geometric phase that arises for half-integer 
values of $S$.} \label{fig.quad_nontrivialpath} \end{figure}

The spectral degeneracy pattern of the quadrumer for half-integer spins is different from that of integer spins. We have seen this distinction earlier in the trimer problem. This suggests a role for the Berry phase term, with a non-zero phase accruing along certain paths in the ground state space. There are several types of closed paths on the quadrumer CGSS consisting of three touching tori. We find that paths within a single torus (with or without winding in either direction) accrue trivial phases that are multiples of $2\pi$. Likewise, paths lying on two tori are also trivial. Non-trivial phases emerge only in paths that traverse all three tori, with a net $\pi$-winding in the vertical direction on each torus. An example is shown in Fig.~\ref{fig.quad_nontrivialpath} (top), consisting of three segments, P-Q, Q-R and R-P, one on each torus. This describes a closed path that crosses from one torus to another at common lines. All three segments correspond to a fixed value of $\phi_1$, so that the first spin remains stationary. Each of the other three spins rotates by $2\pi$, subtending an area of $2\pi$ at the north pole. This corresponds to a net Berry phase of $6\pi S$. For integer spins, this is a trivial phase as it is a multiple of $2\pi$. However, for half-integer spins, we have a physically relevant $\pi$ phase. 

In the `particle in the CGSS' description, this can be understood as an Aharonov-Bohm phase. It corresponds to two $\pi$-flux tubes threaded in the space between tori, as shown in Fig.~\ref{fig.quad_nontrivialpath} (bottom). As seen from the figure, a closed loop on any one torus does not incur a net phase; e.g., a path along the outer torus encloses a net flux of $2\pi$, equivalent to no flux at all. The only paths that are sensitive to the fluxes lie on all three tori, effectively traversing half of each torus.

\begin{figure}
\includegraphics[width=\columnwidth]{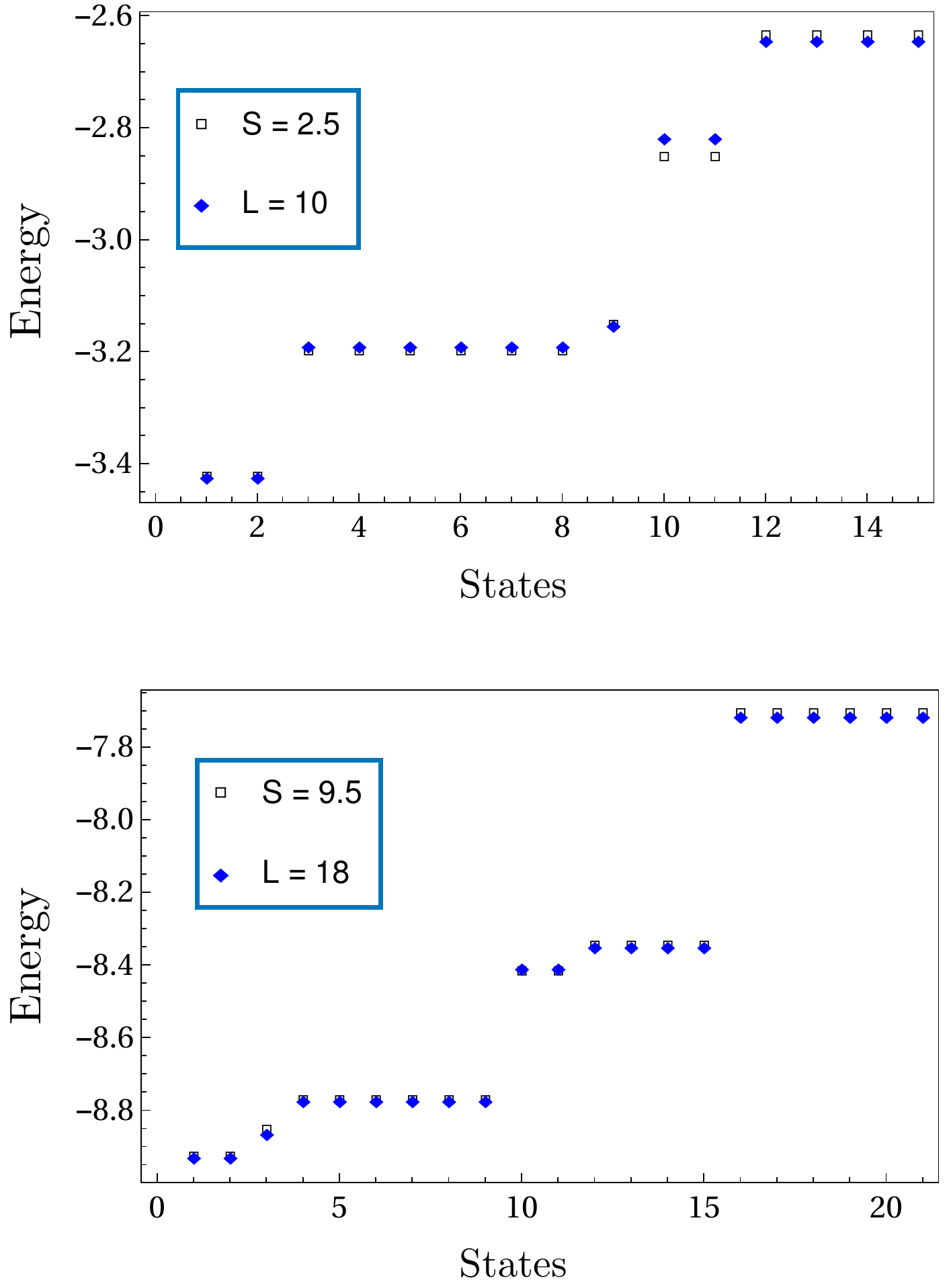}
\caption{Comparison of spin (empty squares) and tight-binding (solid diamonds) spectra for half-integer $S$. Spin spectrum is shifted in order to achieve the same ground state energy as that of the tight-binding spectrum. }
\label{fig.tb_hintS} \end{figure}

We can modify our earlier tight-binding prescription from Sec.~\ref{ssec.quad_tb_int} to take these flux tubes into account. The flux tubes lead to a vector potential on the torus surfaces. This adds a complex phase to the hopping amplitudes, via the well-known Peierls' substitution prescription~\cite{Peierls1933}. 
We assume a simple form of the vector potential that gives rise to the required Aharonov-Bohm phase. We take it to be non-zero on one torus alone, say the torus on the left in Fig.~\ref{fig.quad_mesh}. We take it to have the form $\vec{A} = 2\pi \hat{y}/L$, pointing in the vertical direction. When the particle goes around this torus in the vertical direction, it gains a phase of $2\pi$. It can be easily be checked that this provides the required Aharonov-Bohm phase. For instance, the non-trivial path shown in Fig.~\ref{fig.quad_nontrivialpath} (top) 
accumulates a $\pi$ phase. We solve this tight-binding model numerically and compare it with the half-integer spin spectrum below.

\subsection{Comparison with full quantum description for half-integer spins}
\label{ssec.quad_qm_hint2}

As with integer spins, we solve the half-integer-spin $XY$ quadrumer problem by exact diagonalization. We use total $S_z$ and cyclic permutation symmetries. The resulting spectrum shows a doubly degenerate ground state, unlike integer spins. We fit the spectrum to the tight-binding model with $\pi$-fluxes, treating $L$ and $t$ as fitting parameters as described in Sec.~\ref{ssec.quad_qm_int}. The results, presented in Table~\ref{tab.spin_halfs}, show excellent quantitative agreement. The number of matched states/levels increases with $S$, indicating that the mapping to the tight-binding model becomes more accurate as $S$ increases. Fig.~\ref{fig.tb_hintS} compares the spectra from exact diagonalization of the
spin system and the tight-binding model for two half-integer spins, 
$S = 2.5$ and $9.5$. 

\begin{table}
\begin{centering} 
\begin{tabular} { | c | c | c | c |} 
 \hline
Spin & $L$ & Hopping & Levels matched \\ 
$S$ & & $t$ & (No. of states) \\
\hline
\hline
~~0.5~~ & ~~6~~ & ~~0.5864~~ & 2 ~(8) \\
1.5 & 10 & 0.83687 & 4 ~(11) \\ 
2.5 & 10 & 0.72868 & 5 ~(15) \\
3.5 & 10 & 0.60952 & 5 ~(15) \\
4.5 & 12 & 0.90943 & 5 ~(15) \\
5.5 & 14 & 1.24024 & 5 ~(15) \\
6.5 & 16 & 1.62682 & 6 ~(21) \\
7.5 & 16 & 1.55958 & 6 ~(21) \\
8.5 & 18 & 1.99109 & 6 ~(21) \\
9.5 & 18 & 1.92926 & 6 ~(21) \\
10.5& 20 & 2.40964 & 6 ~(21) \\
11.5& 22 & 2.92317 & 8 ~(27) \\
\hline
\end{tabular}

\caption{ Comparing spin and tight-binding spectra for half-integer $S$ values. $L$ and $t$ parameters shown are obtained by our fitting procedure. The last column shows the number of low-energy levels (and states) that match in the two approaches.}
\label{tab.spin_halfs}
\end{centering}
\end{table}

\section{Does order by disorder determine the quadrumer spectrum?}
\label{sec.ObDquad}

The CGSS for the symmetric quadrumer is much larger than the symmetries of the problem, indicating an accidental degeneracy. 
We may expect low-energy states to sample a `selected' subset of the CGSS. Indeed, this is consistent with our numerical results. For instance, the spectrum for $S=12$ shown in Fig.~\ref{fig.tb_intS} resembles that of a particle on three disjoint rings. The ground state is nearly three-fold degenerate, while excited states are approximately sixfold degenerate.
In addition, the energies vary as $\sim n^2$, where $n$ is an integer. This is in reasonably good agreement with the spectrum of a particle on three disjoint rings $(S^1 \otimes S^1 \otimes S^1)$. This pattern appears for half-integer spins as well, as seen in Fig.~\ref{fig.tb_hintS} for $S=9.5$. We may deduce that the particle is localized around the three collinear lines in the CGSS. Naively, this appears to be consistent with order by disorder as quantum fluctuations tend to favor collinear states over coplanar states. We argue that this is \textit{not} the case. Rather, a more subtle mechanism operates to select collinear lines.

\begin{figure}
\includegraphics[width=\columnwidth]{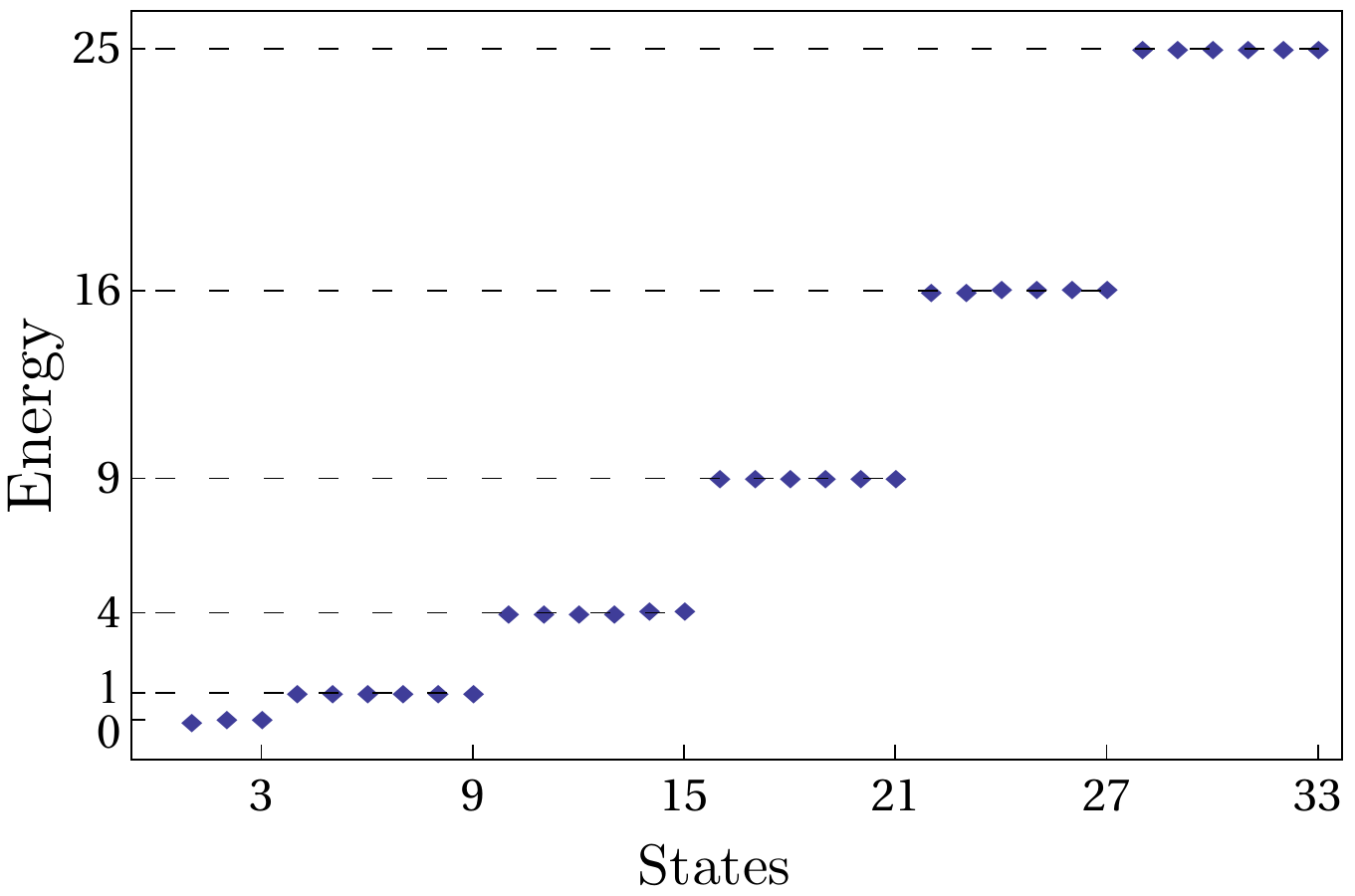}
\caption{Symmetric quadrumer spectrum for $S=19$. The energies have been 
shifted downwards by $-737.686 J$ and scaled by 0.129083. The spectrum 
closely resembles that of a particle on three disjoint rings.} 
\label{fig.S19} \end{figure}

To argue against order by disorder, we compare the selection effect for different $S$ values. Figure~\ref{fig.S19} shows the spectrum for a large spin value, $S=19$, obtained by exact diagonalization. This shows near perfect agreement with the picture of a particle on three disjoint circles. Comparing Fig.~\ref{fig.tb_intS} and Fig.~\ref{fig.S19}, we see that the selection effect becomes stronger with increasing $S$. While order by disorder vanishes in the classical limit, our results suggest that the selection effect is strongest when $S\rightarrow\infty$. This cannot be the result of a $1/S$ correction term as stipulated by the order by disorder paradigm. 

\subsection{Holstein-Primakoff analysis}
\label{ssec.HPquad}

We now present a HP analysis to examine how order by disorder 
may be induced by quantum fluctuations; we will then argue that order
by disorder does not play a role in this problem.
The symmetric quadrumer is a special case of the
asymmetric quadrumer. We can adapt the HP analysis of Sec.~\ref{ssec.asym_quad_HP} to this case by setting $\lm = 0$. We choose the reference state to have $\vec{S}_1=-\vec{S}_2$ and $\vec{S}_3=-\vec{S}_4$, with the state described by two angles, $\phi_1$ and $\phi_{31}$ (the angular distance between the third and first spins). 

Substituting $\lm =0$ in Eqs.~\eqref{ham43}
and \eqref{ombc}, we again find two free particles and two SHOs for a
generic value of $\phi_{31}$. The ground state energy is given by 
\bea E_0 = - 2J S^2 - 2 J S + JS \Big[\cos (\frac{\phi_{31}}{2}) + \sin 
(\frac{\phi_{31}}{2})\Big]. ~~\label{e51} \eea
As in the asymmetric case, this has minima at $\phi_{31}= 0$ and $\pi$. 

If we assume that order by disorder occurs and thereby set $\phi_{31}= 0$, we 
find that the frequency of the SHO corresponding to $(p_c,x_c)$ vanishes. This
is in contrast to the asymmetric quadrumer where the two SHO frequencies take non-zero values for any $\phi_{31}$. Here, we obtain three free particles and one SHO 
with frequency $2JS$, corresponding to $(p_b,x_b)$. This is a manifestation of the non-manifold character of the CGSS. At generic points, it is two-dimensional (with two free particles in the HP description). In contrast, at collinear states where two tori touch, we have additional degrees of freedom allowing for motion onto a different torus. This is reflected as an additional free particle in the HP analysis.

Taking the HP result at face value, we may expect collinear 
states to be selected with the lowest branch of excitations corresponding to
\beq H_{free} = \frac{JS}{2} (P^2 + p_a^2 + p_c^2), \label{ham52} \eeq
corresponding to 
a particle moving on a three-dimensional torus, $T^3$. We note that the $p_a$ and $p_c$ take the system away from collinearity. Treating them as free particles is not consistent with our assumption of order by disorder. It is conceivable that higher order terms will introduce confining potential energy terms in $x_a$ and $x_c$. 
We may expect 
to see the spectrum of a particle on three disjoint circles, with three-fold degeneracy arising from the three possible ways of choosing a collinear configuration. Notably, as the zero point energy is a $1/S$ effect, the selection effect should weaken as $S$ increases (see the
discussion of the asymmetric quadrumer in Sec.~\ref{ssec.asym_quad_HP}). However, this is not consistent with our numerical results which show stronger selection for larger $S$. 

A second piece of evidence against order by disorder comes from the nature of our tight-binding model. We find excellent agreement between the tight-binding spectrum and exact diagonalization, with the agreement improving with increasing $S$. This agreement is achieved without including a potential-like term that would arise from Eq.~\eqref{e51}. Apart from hopping between nodes, the particle would experience a local potential which has minima at collinear lines. 
The irrelevance of this zero point potential energy indicates that order by disorder is not applicable here.

While the HP approach does not explain the full spectrum (as compared to the tight-binding model), we note that it brings out the non-manifold character of the CGSS, with a SHO turning into a free particle for collinear reference states.

\section{Order by singularity: evidence from exact diagonalization}
\label{sec.ObSED}

Our numerical results show that at large $S$, the symmetric quadrumer
resembles a particle moving on three disjoint circles. In the above discussion, we have surmised that this indicates selection of collinear states over others within the CGSS. We demonstrate here that: (a) such selection does not occur in the asymmetric quadrumer which has a 2D manifold as CGSS, (b) the symmetric quadrumer, with its non-manifold CGSS, shows a preference for collinear states even in the $S\rightarrow\infty$ limit. We provide two pieces of evidence from the numerical solution of the spin problem using exact diagonalization.

\subsection{Measuring collinearity}

We consider a diagnostic operator of the form
\begin{eqnarray} \non
\hat{O}_{coll.} &=& \left(({\vec{S}}_1\cdot {\vec{S}}_2 )_H({\vec{S}}_3\cdot 
{\vec{S}}_4 )_H + ({\vec{S}}_1\cdot {\vec{S}}_3 )_H({\vec{S}}_2\cdot 
{\vec{S}}_4 )_H\right.\\
&+& \left.({\vec{S}}_1\cdot {\vec{S}}_4 )_H({\vec{S}}_2\cdot {\vec{S}}_3 )_H
\right)/(S(S+1))^2. \label{eq.Odef} \end{eqnarray}
Here, $(\vec{S}_i \cdot \vec{S}_j)_H \equiv S_i^x S_j^x + S_i^y S_j^y + 
S_i^z S_j^z$, a Heisenberg dot product. 
We call this operator $\hat{O}_{coll.}$ as it provides a measure of collinearity as discussed below.
All the terms in Eq.~\eqref{eq.Odef} are quartic in spin operators. We empirically find that scaling by $S^2(S+1)^2$, rather than $S^4$, allows for a smooth fit as a function of $S$.
For a given $S$, we calculate the `quantum' expectation value of this operator in the numerically obtained ground state. We also find its `classical expectation value', defined as follows. This operator is evaluated in each classical ground state by replacing spin operators with the corresponding classical vectors. This result is averaged over all classical ground states, i.e., all points in 
the CGSS. We compare the quantum and classical expectation values. We may naively expect these two to coincide in the semiclassical limit by the following argument. In the `particle in the CGSS' picture, at low energies, we expect the particle to sample all points in the CGSS equally. This can also be argued from a path integral based evaluation of expectation values, with all classical ground states contributing with equal weight. The quantum expectation value must be the average over all points in the CGSS.

\begin{figure}
\includegraphics[width=\columnwidth]{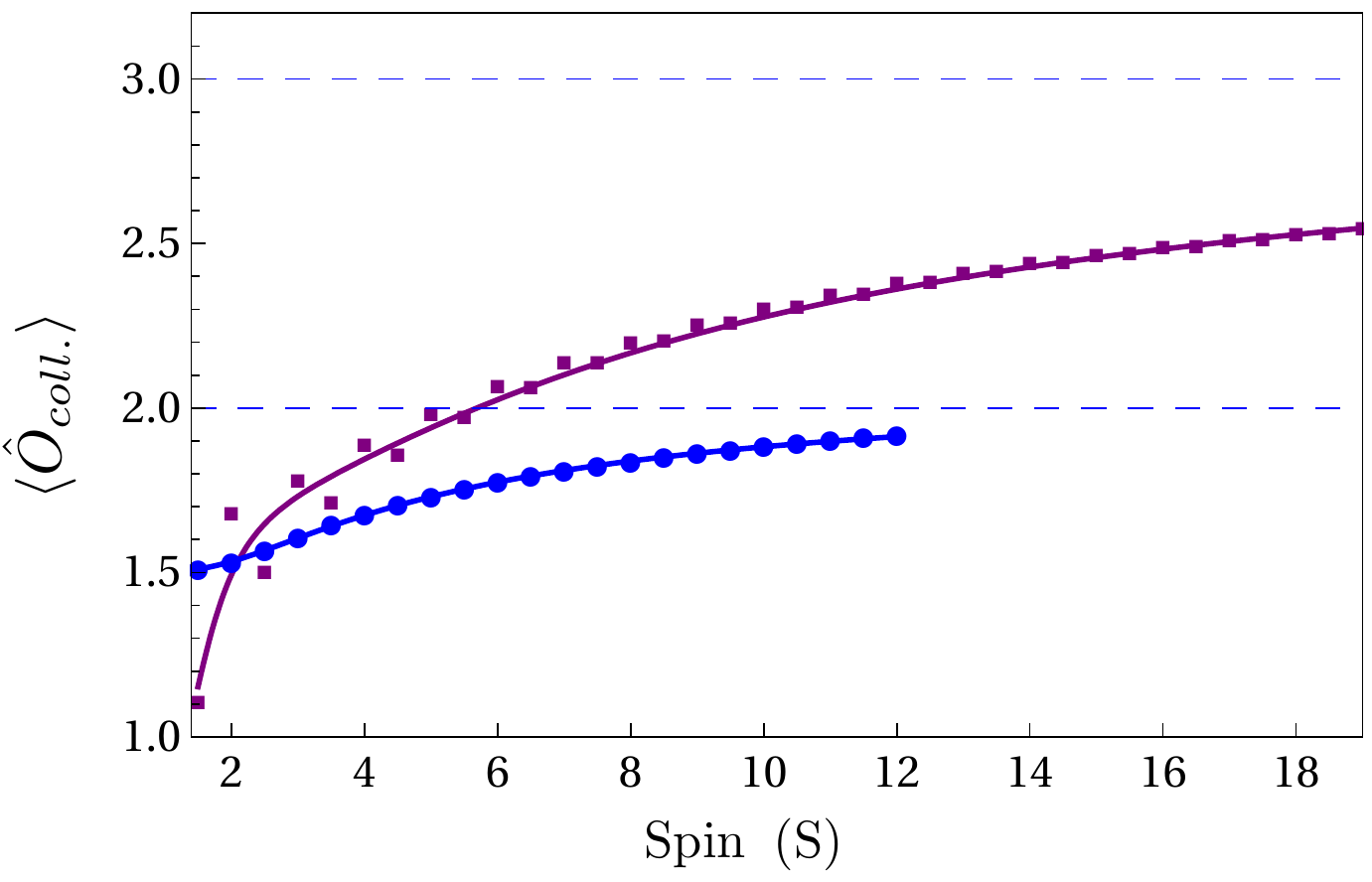}
\caption{Expectation value of $\hat{O}_{coll.}$ vs $S$. Data for the asymmetric quadrumer is shown as blue dots. The corresponding fitting curve (blue line) 
is $f_{\mathrm{asym}}(S) = 2.09584 - 2.48505/S + 3.654/S^2 - 1.87011/S^3$.
Data for the symmetric quadrumer is shown using violet squares. The fitting 
function (violet line) is given by $f_{\mathrm{sym}}(S) = 2.96877 - 9.45947/S + 29.6598/S^2 - 45.1815/S^3 + 23.7693/S^4$.} \label{fig.Ocoll} \end{figure}

Figure~\ref{fig.Ocoll} shows the quantum expectation value of $\hat{O}_{coll.}$ vs $S$ for two problems: the asymmetric quadrumer with $\lm=2$ and the symmetric quadrumer ($\lm=0$). In the former, we see that $\la \hat{O}_{coll.} ^{a.quad} \ra_{quantum} \rightarrow 2$ as $S\rightarrow \infty$ (see the fitting function in 
the caption of Fig.~\ref{fig.Ocoll}). To obtain the classical expectation value, we note that the CGSS for the asymmetric quadrumer is a single torus as shown in Fig.~\ref{fig.aquadgs}. In terms of $\phi_1$ and $\phi_3$, we find 
\bea \la {O}_{coll.}^{a.quad.} \ra_{class.}(\phi_1,\phi_3) ~=~ 1 ~+~ 2 \cos^2 
(\phi_{3}-\phi_1). \eea
The first term is unity, as $({\vec{S}}_1\cdot {\vec{S}}_2 )_H/
(S(S+1))$ and $({\vec{S}}_3\cdot {\vec{S}}_4 )_H/(S(S+1))$ both 
take the value $-1$ in all the
classical ground states. The other two terms evaluate to $\cos^2 (\phi_{3}-\phi_1)$. In order to average over the CGSS, we average over all values of 
$\phi_3-\phi_1$. This gives $\la {O}_{coll.}^{a.quad} \ra_{CGSS} = 2$. Our numerical result for the quantum expectation value coincides with this value as shown in Fig.~\ref{fig.Ocoll}. This provides evidence that the asymmetric quadrumer maps to a particle that samples every point on the CGSS. 

For the symmetric quadrumer, the CGSS consists of three copies of the 
asymmetric quadrumer CGSS. The classical expectation value on each torus has the same form as that for the asymmetric quadrumer given above. Averaging over the three tori, we expect to find $\la {O}_{coll.}^{s.quad} \ra_{CGSS} = 2 $. However, this does \textit{not} agree with the numerically obtained quantum expectation value. As seen in Fig.~\ref{fig.Ocoll}, the latter extrapolates to $\la \hat{O}_{coll.} ^{s.quad} \ra_{quantum} \rightarrow 3$ as $S\rightarrow \infty$ (see fitting curve). Notably, this saturates an upper bound, being the maximum possible classical value for $\hat{O}_{coll.}$. This maximum value is only reached in collinear states where each term in $\hat{O}_{coll.}$ gives 1. In the `particle in the CGSS' picture for $S\rightarrow \infty$, we conclude that the particle only samples collinear states and not the entire CGSS. 

\subsection{Spin correlations}

The quantum ground state wave function contains information about spin correlations. However, a direct evaluation of correlations functions cannot distinguish between uniform sampling on the CGSS and the
selection of collinear states. We have devised the following diagnostic that we apply to the asymmetric and symmetric quadrumers. 

\begin{figure}
\includegraphics[width=\columnwidth]{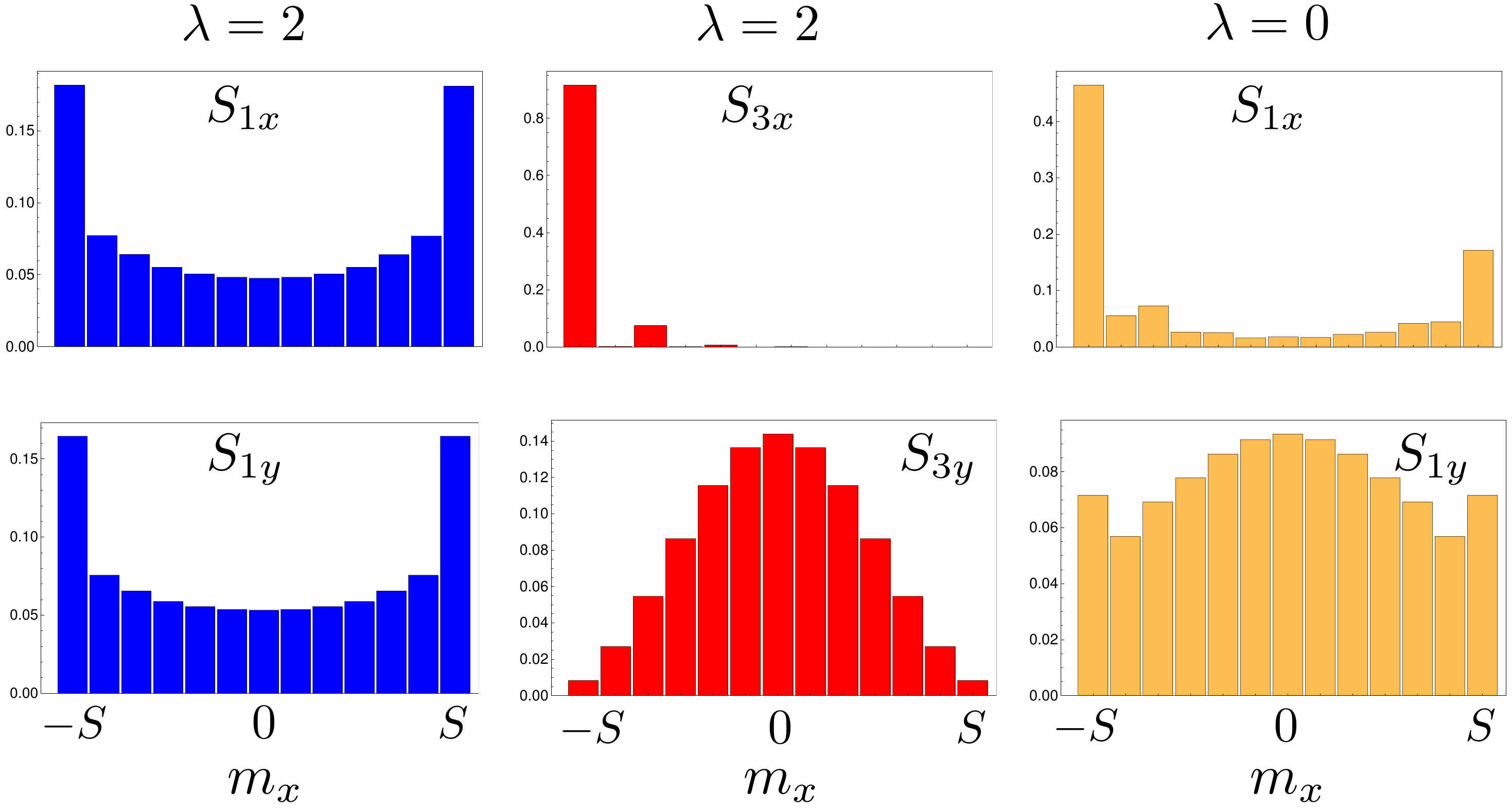}
\caption{Probability distributions in the numerical ground state projected to fix $S_{4}^x = S$. See text
for details.} \label{fig.Projs} \end{figure}

We take the numerically obtained ground state in the $S_z$ basis, given by
\begin{equation} \vert GS\ra = \sum_{m_1, m_2,m_3,m_4} a_{m_1, m_2,m_3,m_4}
\vert m_1, m_2,m_3,m_4\ra, \end{equation}
where $\hat{S}_i^z \vert m_i\ra = m_i \vert m_i\ra$.
We act with a projection operator $\hat{P}_{4,x}$ on this state to project the fourth spin $\vec{S}_4$ along the $\hat{x}$ direction, i.e., we pick out the component of the ground state with $S_{4}^x=S$. After normalization, this gives
\bea \nonumber \vert GS\ra_{proj.} = \frac{\hat{P}_{4,x} \vert GS\ra}{\vert \la GS\vert
\hat{P}_{4,x} \vert GS\ra \vert^{1/2}} = \frac{\la S_4, S_{4x} = S\vert GS\ra}{\vert\vert
\la S_4, S_{4x} = S\vert GS \ra\vert\vert}. \phantom{ab}\eea 
We write this as
\begin{equation} \vert GS\ra_{proj.} = \sum_{m_1, m_2,m_3} b_{m_1, m_2,m_3}
\vert m_1, m_2,m_3\ra. \end{equation}
We now note that any $\hat{S}_{j}^{x/y}$ operator eigenstate can be written as a linear combination of $\hat{S}_{j}^{z}$ eigenstates,
\begin{equation} \vert S_j, S_j^{x/y} = \mu\ra = \sum_{m'_j} c^{x/y}_{\mu,m'_j}
\vert m'_j\ra. \end{equation} 
We resolve the (projected and normalized) wave function $\vert GS\ra_{proj.}$ into different $S_{j}^{x/y}$ components. For example, an $S_{1}^{x/y}$ component is given by
\begin{eqnarray} &&\la S_1, S_1^{x/y} = \mu\vert GS\ra_{proj.} \non \\
&=& \sum_{m_2,m_3}\left(\sum_{m_1} (c^{x/y}_{\mu,m_1})^*\phantom{a} b_{m_1, 
m_2,m_3}\right) \vert m_2,m_3\ra. \end{eqnarray}
We deduce that the probability of having $S_{1}^{x/y} = \mu$ is 
\begin{equation} \sum_{m_2,m_3}\left|\left(\sum_{m_1} (c^{x/y}_{\mu,m_1})^*
\phantom{a} b_{m_1, m_2,m_3}\right)\right|^2. \end{equation}
We find these probabilities for different spin components.
Figure~\ref{fig.Projs} shows the resulting probability weights for $S=6$. 
We interpret the results as follows. 

Figure~\ref{fig.Projs} (left and center) show the results for the asymmetric quadrumer. In this problem, $\vec{S}_3$ and $\vec{S}_4$ are strongly antiferromagnetically coupled. As a result, $\vec{S}_3$ is anti-aligned with $\vec{S}_4$ in the classical ground state. However, $\vec{S}_1$ does not have a fixed orientation with respect to $\vec{S}_4$; the CGSS includes states with all possible relative orientations between $\vec{S}_1$ and $\vec{S}_4$. This is reflected in Fig.~\ref{fig.Projs}, where $\vec{S}_4$ has been fixed along the $\hat{x}$ direction by a projection operation. The probability weights for ${S}_1^{x/y}$ are shown in the
left panels. We see peaks at ${S}_1^{x}=\pm S$ and ${S}_1^{y}=\pm S$. This is consistent with 
the spins lying in the $x-y$ plane with no preferred orientation; if we consider a semiclassical picture
with $S_1^x = S \cos \phi$ and assume that all values of $\phi$ are equally
likely, then the probability distribution of $S_1^x$ would be given by
\bea P(S_1^x) &=& \int_0^{2\pi} ~\frac{d\phi}{2\pi} ~\de (S_1^x ~-~ S \cos 
\phi) \non \\
&\sim& \frac{1}{\sqrt{S^2 ~-~ (S_1^x)^2}}, \eea
which has peaks at $S_1^x \simeq \pm S$. (A similar argument works for
$S_1^y$). In contrast, the middle panels in Fig.~\ref{fig.Projs} show 
that the probability weight for ${S}_3^{x}$ is sharply peaked at $m_x=-S$ while the probability weight for ${S}_3^{y}$ 
does not indicate a preference for direction. Taken together, they indicate that $\vec{S}_3$ is pinned along the $-\hat{x}$ direction. This allows for an elegant interpretation in the semiclassical limit: the quantum ground state can be thought of a uniform sampling of the CGSS. In other words, the particle on the CGSS has a ground state wave function that is uniformly weighted on the space. 

The panels on the right in Fig.~\ref{fig.Projs} show the results for the symmetric quadrumer. As above, we have projected the wave function to fix $\vec{S}_4$ along the $\hat{x}$ direction. We only show the probability weights for ${S}_1^{x/y}$ as $\vec{S}_2$ and $\vec{S}_3$ show the same results due to symmetry. Remarkably, the probability weight is peaked at $S_1^x=-S$, with a sub-dominant peak at $S_1^x=S$. Semiclassically, this can be understood as arising from the average over collinear states. We note that there are three distinct collinear states. One collinear state has $\vec{S}_1$ parallel to $\vec{S}_4$ while two have it anti-aligned. Averaging over these three, we expect $S_1^x=-S$ to have a probability weight of $\approx 0.66$, while $S_1^x=S$ should have $\approx 0.33$. Our numerical results are close to this
expectation, in that the ratio of the two probability weights is close to 2.
The agreement may improve for larger values of $S$. 

\section{Order by singularity: insight from the tight-binding model}
\label{sec.ObSTB}

The mechanism behind the selection of collinear states is best understood from the tight-binding model. In Figs.~\ref{fig.tb_intS} and \ref{fig.tb_hintS}, we demonstrated that the tight-binding model provides excellent quantitative agreement with the spectrum. We have also shown that the agreement improves as $S$ increases. On the strength of this agreement, we take the tight-binding wave functions to be an accurate representation of the spin states.

\begin{figure}
\includegraphics[width=0.7\columnwidth]{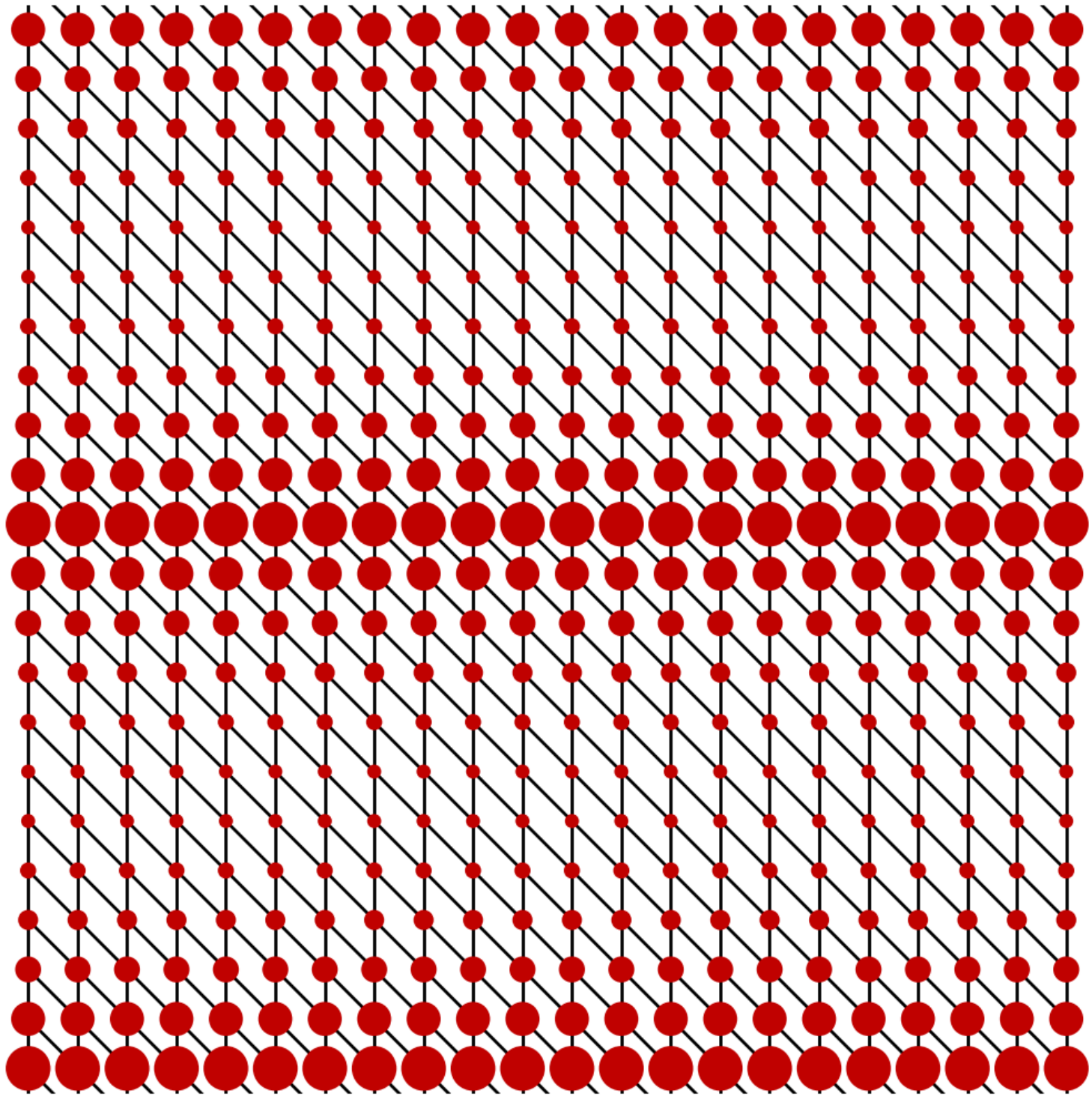}
\caption{Ground state obtained using the tight-binding approach. The tight-binding mesh has $22\times 22$ sites on each torus with common lines being 
identified. The size of the circle at each point is proportional to the local probability 
density in the ground state wave function.} \label{fig.TBgs} \end{figure}

The wave functions obtained from the tight-binding model clearly reveal the mechanism underlying the selection of collinear states. We recapitulate that the space consists of three tori that touch along singular lines. Remarkably, we find that all low-lying wave functions are localized with dominant weight around the singular lines. Figure~\ref{fig.TBgs} shows the probability weights extracted from the tight-binding ground state. The wave function is symmetric among the 
different tori. Hence the figure shows only a single torus, as the same weights are repeated on the other two tori. Surprisingly, the weights are sharply peaked on the common singular lines. We find the same localization pattern in all low-lying states. Below, we explain this observation using an analytic study of bound states in the tight-binding model. 

\subsection{Bound state wave functions}
\label{ssec.boundstates}

\begin{figure}
\includegraphics[width=\columnwidth]{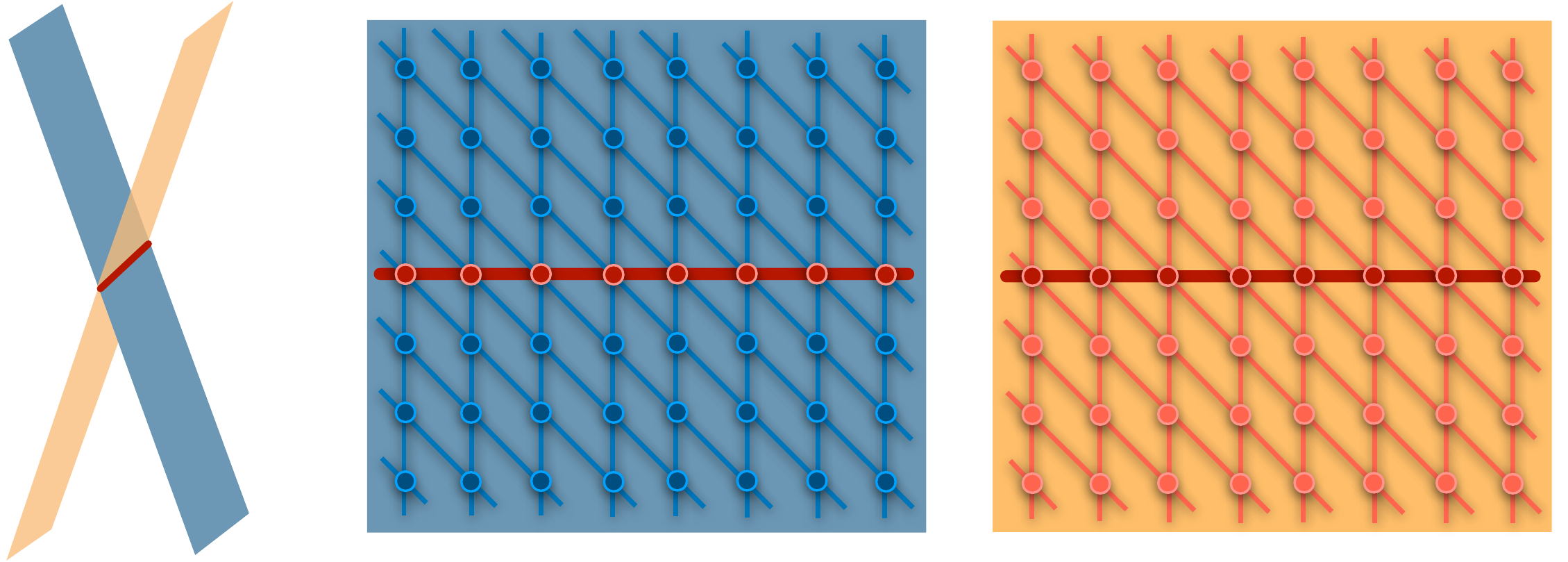}
\caption{Set up for calculating bound states in the tight-binding model. Left: A zoomed out view of two sheets intersecting along a line. We assume periodic boundary conditions with the intersection line closing on itself to form a circle. Center and Right: Tight-binding mesh on two sheets that share a line of points. We have diagonal and vertical bonds. Note that there are no horizontal bonds, either along 
the central line or elsewhere. 
 A generic site is connected to four nearest neighbors. However, sites on the 
common line are connected to eight neighbors, four on each sheet.}
\label{fig.TB_bound_state_set_up} \end{figure}

We consider the tight-binding model in the limit of large system size, $L$. The set up is shown in Fig.~\ref{fig.TB_bound_state_set_up}, with two sheets intersecting along a line. We are interested in bound states localized along this singular line. The sheets themselves are tori with periodic boundaries. For large $L$ and sharply localized bound states, we may ignore the periodicity and work with open sheets.

The tight-binding Hamiltonian is given by 
\beq H ~=~ -t ~\sum_{m} ~\sum_{n(m)} ~c_{m}^\da c_{n}, \eeq
where the index $m$ runs over all sites in our mesh over the two sheets 
shown in Fig.~\ref{fig.TB_bound_state_set_up}. The sum over $n(m)$ represents a sum over sites that are connected to $m$ by a bond. A generic point has neighbors within the same sheet. However, points on the singular lines have neighbors on two tori. The hopping amplitude is a constant, $-t$. In this single-particle Hamiltonian written in the site-basis, an eigenfunction is given by $\vert \psi \rangle \equiv \sum_m \psi_m \vert m \rangle$, where $\vert m \rangle$ denotes a 
state localized at site $m$ which is occupied with amplitude $\psi_m$.
In this language, an eigenstate with eigenvalue $E$ satisfies 
\beq -t ~\sum_{n(m)} ~\psi_n ~=~ E \psi_{m}. \label{eq.evaleq} \eeq

We first consider a non-localized state that is not bound to the
singular lines. Such a state is largely weighted away from the collinear lines, which are a 1D subset of the full 2D space. Away from the collinear line, the space looks very much like two independent sheets. A particle freely moving on this space can be thought of as having eigenstates characterized by 
two momenta, $k_x$ and $k_y$. The energy is given by 
$E_{tor} = -2t \left[ \cos(k_y) + \cos(k_y-k_x)\right]$. As $k_x$ and $k_y$ can take any value between $0$ and $2\pi$ (assuming periodic boundaries), the energy falls within a range, $E_{tor} \in [-4t,4t]$. Below, we will consider 
an ansatz for the bound states. If a candidate bound state has energy lying within the range $[-4t,4t]$, it will hybridize with the `free' states that are not localized. Thus, it is unlikely to be bound. However, if we find a candidate state with energy below $-4t$, it will remain localized.

We consider a bound state ansatz given by
\begin{eqnarray} \psi_{p,q,T} = \left\{ 
\begin{array}{cc}
e^{i 2\pi \ell p/L}, & q=0 \\
e^{ -\al \vert q \vert } e^{i 2\pi \ell (p+q/2)/L} , & q \neq 0 
\end{array} \right.. \end{eqnarray}
Here, $T=1,2$ represents the two sheets in the problem. The indices
$p$ and $q$ are integers that label sites on each sheet, in the horizontal and vertical directions respectively. In particular, $q=0$ corresponds to the 
line shared between the two sheets. Its horizontal extent is assumed to be $L$. On this line, the $T$ index loses its meaning as the sites are shared by both sheets.
By construction, the ansatz wave function is symmetric between the two sheets and is localized along the shared line, decaying exponentially as we move away from it. We have introduced a parameter $\ell \in \mathbb{Z}$ (to guarantee periodicity along the horizontal direction). This represents an angular momentum index. It indicates the degree of phase winding as we move along the line. 

We now consider $(p,q,1)$ with $q\neq 0$, i.e., a site that is not on the shared line. It has four neighbors, given by $(p,q-1,1)$, $(p,q+1,1)$, $(p+1,q-1,1)$ and $(p-1,q+1,1)$. The eigenvalue equation Eq.~\eqref{eq.evaleq} with reference to this site, after a few simplifications, gives 
\begin{eqnarray} E ~=~ -2t ~[e^\al + e^{-\al}]
~\cos \left(\frac{\pi\ell}{L} \right). \label{eq.E2l} \end{eqnarray}
We obtain the same equation from a generic point on the second sheet ($T=2$) as well, providing a consistency check. 

We now consider a site on the line, $(p,0,T=1/2)$. This site has eight neighbors: $(p,-1,T)$, $(p,+1,T)$, $(p+1,-1,T)$ and $(p-1,+1,T)$, with $T=1,2$. The eigenvalue equation with reference to this site gives
\begin{eqnarray} E ~=~ -8t ~e^{-\al} \cos \left(\frac{\pi \ell}{L} \right).
\label{eq.E3l} \end{eqnarray}
Comparing Eqs.~\eqref{eq.E2l} and \ref{eq.E3l}, we obtain 
\begin{eqnarray} \al ~=~ \frac{\ln{3}}{2}. \end{eqnarray}
This fixes the decay length in the bound state ansatz. Remarkably, we find the same decay length for any value of $\ell$. From Eq.~\eqref{eq.E3l}, we 
obtain the energy, 
\begin{equation} E ~=~ - ~\frac{8t}{\sqrt 3}~ \cos \left( \frac{\pi\ell}{L}
\right) ~\simeq~ -4.6188 t ~\cos \left( \frac{\pi\ell}{L} \right). 
\label{eq.Eell} \end{equation}
The energy naturally depends on $\ell$, the angular momentum quantum number. The lowest energy occurs for $\ell = 0$, giving rise to a real wave function without any phase winding. As $\ell$ is increased from zero, the energy increases. 

We have identified bound state solutions. However these represent 
true bound states only if their energies lie outside the range of energies of 
the delocalized states. We define a critical angular momentum, $\ell_c$, where the bound state energy enters the delocalized continuum. We then obtain
\begin{eqnarray} - ~\frac{8t}{\sqrt 3} ~\cos \left( \frac{\pi\ell_c}{L} \right)
~\simeq~ -4t ~\implies~ \ell_c ~\simeq~ \frac{L}{6}. \end{eqnarray}
This is a remarkable result that indicates that we have true bound states for $\ell = 0, \pm 1, \pm 2, ...,\pm L/6$. In other words, we have about 
$L/3$ true bound states localized along the shared line. 

In the full CGSS of the symmetric quadrumer, we have three distinct singular lines. Each of them can host bound states independently. When $L$ is not too small, the bound states on one line decay well before a second line is approached. This indicates that the
bound states do not hybridize. As we have three shared lines and $\sim L/3$ bound states per line, we have $\sim L$ bound states in the system, a macroscopic number. We conclude that the low-energy spectrum consists solely of bound states, with the number of bound states scaling as the linear size of the system. 

The requirement of large $L$ corresponds to large values of $S$ in the quantum 
problem (see Tables~\ref{tab.intspin} and \ref{tab.spin_halfs}). For smaller 
values of $S$, we have a tight-binding problem with a small $L$. This leads to hybridization between bound states on different shared lines. This is responsible for the degeneracy pattern seen in the tight-binding dispersion. For instance, for $S=19$ in Fig.~\ref{fig.S19}, we find a nearly three-fold degenerate ground state. For smaller (integer) $S$ values shown in Fig.~\ref{fig.tb_intS}, this is broken down into a non-degenerate ground state and two excited states. 
 
Having established that the low-energy states of the tight-binding model are all bound to singular lines, we revisit the conjecture described in Sec.~\ref{sec.quad}, viz., that the low-energy physics of the symmetric quadrumer maps to a single particle moving on its CGSS. 
We have subsequently shown that the tight-binding model faithfully reproduces the low-energy exact diagonalization spectrum of the quadrumer. If the agreement were only restricted to bound states, this would cast doubt on the tight-binding model as a true effective model. For example, it could be construed that some selection mechanism (perhaps a stronger version of order by disorder) operates to pick collinear states. This binds low-energy states to the collinear lines, with tunnelling processes on the surface of the tori. However, a closer examination of our results allows us to refute this contention. As seen from Tables~\ref{tab.intspin} and \ref{tab.spin_halfs}, the number of matching states (when comparing the tight-binding and exact diagonalization spectra) is always larger than $L$. As $L$ is the number of tight-binding-bound-states, we see that the agreement between the models extends to unbound, delocalized states as well. 
We also see this directly from the spatial profiles of the matching tight-binding eigenfunctions. This gives us confidence that the tight-binding model on the CGSS indeed provides a true effective theory of the symmetric quadrumer.

\section{Summary and Discussion}
\label{sec.summary}

\renewcommand{\arraystretch}{1.3}
\begin{table*}
\begin{tabular}{|c|c|c|c|}
\hline
& \multicolumn{2}{c|}{Space of effective dynamics} & Nature \\ \cline{2-3}
& Integer $S$ & Half-integer $S$ & of CGSS \\ \hline
Dimer &\multicolumn{2}{c|}{A ring ($S^1$)} & 1D manifold \\ \hline
Trimer & Two disjoint & Two disjoint rings & 2 copies of a\\ 
& rings $(S^1\otimes \mathbb{Z}_2)$ & with $\pi$-fluxes threaded 
& 1D manifold\\ \hline
Asymmetric & \multicolumn{2}{c|}{A torus $(T^2)$} & 2D manifold \\
quadrumer & \multicolumn{2}{c|}{}& \\ \hline
Symmetric & Three tori touching& Three touching tori with& Non- \\ 
quadrumer & along lines & two $\pi$-flux tubes threaded& manifold\\ \hline
\end{tabular}
\caption{Summary of results for the various clusters studied here. In each case, a particle moving on the `space of effective dynamics' provides an effective description of the low-energy states.}
\label{tab.summary}
 \end{table*}

We have discussed a paradigm for finding low-energy
effective theories of quantum spin clusters. The interacting spin problem maps to a single particle moving on the space of classical ground states. We have established this equivalence in qualitative and quantitative terms, 
using various clusters with $XY$ antiferromagnetic bonds as test cases. Table~\ref{tab.summary} presents a summary of our results. Geometric phases can play an important role, appearing as Aharonov-Bohm fluxes seen by the particle. 
Using this paradigm, magnetic clusters can be viewed as realizations of toy quantum models. 
This suggests a new route to test theoretical ideas in contexts such as driven rotors\cite{Lin2014}, coupled rotors\cite{Notarnicola2018}, and
rotors with Aharonov-Bohm fluxes\cite{Tian_2010}. 
Our results also serve as a starting point for the study of frustrated molecular and lattice magnets. Our paradigm can be tested in larger magnetic clusters where tools such as the irreducible tensor operator method can be used to evaluate the spectrum\cite{Bencini1990,Schnalle2010}. This could be compared with the appropriate single particle problem. 
Among lattice systems, Er$_2$Ti$_2$O$_7$ is a pyrochlore antiferromagnet with $U(1)$, a circle, as its classical ground state 
space~\cite{Savary2012}. This is analogous to the dimer problem that we have discussed above.

We have proposed a new selection phenomenon, `order by singularity', a consequence of non-manifold structure, wherein the low-energy spectrum consists exclusively of bound states localized around singularities. Perhaps, the most significant aspect of this new mechanism is that it is strongest in the classical $S\rightarrow \infty$ limit. In this light, it provides a counterpoint to early studies of the symmetric quadrumer by Chalker and Moessner\cite{Moessner1998,Chalker2011}. Working with the classical $XY$ quadrumer at low temperatures, they showed that thermal fluctuations `select' collinear states. The selection is sharp with a $\delta$-function-like effective probability distribution. Our results offer a quantum analogue with collinear states being sharply selected by quantum fluctuations. As this selection is strongest at $S\rightarrow \infty$, it is plausible that this has consequences for the purely classical model as well. This is an interesting direction for future studies. We present a plausibility argument in Appendix~\ref{App.class_quad}. We show that the delta-function-like selection effect in the classical model disappears when the quadrumer is made asymmetric, as in Sec.~\ref{sec.asym_quad} above. This removes the singularities in the CGSS. As it also kills the sharp selection effect, it is plausible that singularities play a role in state selection.

Our study has strong parallels with the notion of spontaneous symmetry breaking. It is well-known that finite systems cannot break symmetries to develop order. Rather, they develop low-lying excitations that form an Anderson tower~\cite{Anderson1952}, characteristic of the space of symmetries that will be broken in the thermodynamic limit. In our language, this constitutes a mapping to a particle moving on the classical ground state space. Our analysis shows that geometric phases may have to be taken into account to obtain a satisfactory description. 
Our analysis also extends this notion to frustrated systems, wherein the space can be larger than the symmetry of the problem. In particular, we find interesting effects that arise when the space has singularities. 

Our analysis of the XY quadrumer resonates strongly with the problem of the Heisenberg quadrumer. Two of the current authors have shown that the Heisenberg quadrumer possesses a non-manifold ground state space~\cite{Khatua2018}. It is generically five-dimensional. However, it has three singular subspaces corresponding to collinear states. At these points, the space appears to be six-dimensional. The current $XY$ problem also has the same flavor with a two-dimensional space and three singular lines, corresponding to collinear states. In fact, the $XY$ CGSS is a slice of the Heisenberg ground state space. Remarkably, in the Heisenberg problem, the low-energy states do not consist of bound states, 
and a good effective description is obtained by neglecting singularities~\cite{Khatua2018}. This suggests that not all non-manifolds can induce bound states. The dimensionality of the space and co-dimensionality of the singularities must play an important role. This opens an exciting direction for future studies. 

The experimental consequences of order by singularity also throw up interesting challenges. In the quadrumer cluster, we have shown that order by singularity is a much stronger effect as compared to order by disorder. The latter only gives rise to small quantitative corrections while the former operates over a large range of values of $S$. 
The irrelevance of order by disorder comes from the finiteness of our cluster. With only four spins, the quantum zero point energy does not differ significantly over the classical ground state space (see the discussion of the asymmetric 
quadrumer above). However, this may change in a macroscopic magnet with a non-manifold ground state space.
The consequences of order by singularity and its competition with order by disorder are interesting open questions. 

Our analysis in the context of quantum magnetism has similarities with studies motivated by non-manifold geometries and black hole horizon states~\cite{Balachandran1993,Govindarajan2011,Govindarajan2016}. These studies identify bound states by suitably defining boundary conditions. Our work provides a realistic example where such bound states dominate the low-energy physics. We have used a tight-binding approach on a non-manifold space, an approach with strong parallels to quantum graph studies\cite{Pauling1936,Kottos1997,Keating2008,Harrison2011,Alexandradinata2018}. 

\acknowledgments We thank R. Shankar (Chennai) and T. R. Govindarajan for 
useful discussions. SK thanks Rakesh Netha and Gaurav Sood for help with 
numerics. DS thanks DST, India for Project No. SR/S2/JCB-44/2010 for 
financial support. SK and RG thank the International Centre for Theoretical Sciences (ICTS), Bengaluru, for hospitality during the 2nd Asia Pacific Workshop on Quantum Magnetism (Code: ICTS/apfm2018/11), where a portion of this work was completed.

\appendix

\section{Distances on quadrumer ground state space}
\label{App.dist_quad_gspace}

We have discussed the ground state spaces of the asymmetric and symmetric quadrumers above. The discussion in the main text brings out the topology or the connectivity of these spaces. Here, we discuss the local structure, or loosely the metric, on this space. 

Consider the CGSS of the symmetric quadrumer as shown in Fig.~\ref{fig.quadgs}. We have three tori that touch along lines. Each torus is described by two coordinates, e.g., the torus on the left is described by $\phi_1$ and $\phi_{31}$. Here, $\phi_1$ corresponds to global in-plane spin rotations, a symmetry of the problem. In contrast, $\phi_{31}=\phi_3-\phi_1$ is invariant under rotations. Given a point on the CGSS, we can make small displacements in both variables. If we change $\phi_1 \rightarrow \phi_1 + \de$ while keeping $\phi_{31}$ fixed, this corresponds to rotating each of the four spins by an angle $\de$. The `total displacement', the sum of displacements of all four spins, is $4\de$.
However, keeping $\phi_1$ fixed and changing $\phi_{31} \rightarrow \phi_{31} + \de$ displaces $\vec{S}_3$ and $\vec{S}_4$ by an angle $\de$. It keeps $\vec{S}_1$ and $\vec{S}_2$ fixed. This corresponds to a shorter total displacement of $2\de$.

Similarly, moving along diagonals, we find that the shortest total displacement occurs when moving in the north-west or south-east direction. This corresponds to changing $\phi_1 \rightarrow \phi_1 - \de$ and $\phi_{31} \rightarrow \phi_{31} + \de$. The total displacement is $2\de$ as $\vec{S}_3$ and $\vec{S}_4$ remain stationary. 

\begin{figure}
\includegraphics[width=0.6\columnwidth]{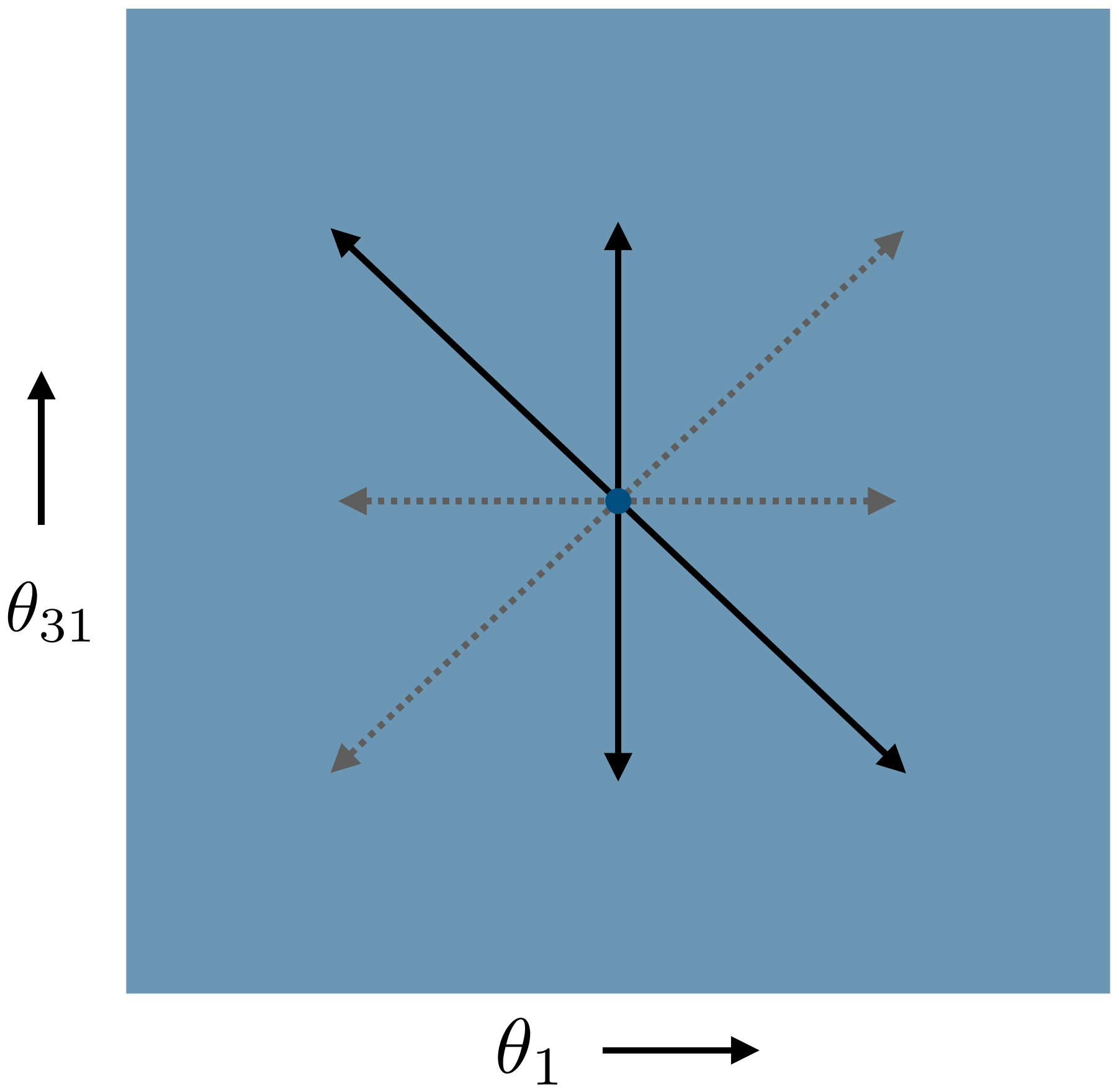}
\caption{Distances on the quadrumer CGSS.} \label{fig.mesh_metric} \end{figure}

Thus, we identify the vertical and north-west/south-east directions as `nearest' distances, as shown in Fig.~\ref{fig.mesh_metric}. Displacement along these directions leads to short total displacements. Moreover, a fixed displacement along either of these directions corresponds to the same total displacement. We use this information to construct a tight-binding model. We include bonds in the vertical and north-west/south-east 
directions with the same hopping amplitude for both.

\section{Holstein-Primakoff theory for the dimer}
\label{App.HPdimer}

We describe the HP analysis of the dimer problem here. 
We note that the Hamiltonian commutes with the $z$-component of the total spin, $S^z \equiv S_1^z + S_2^z$. The eigenvalues of $S^z$, denoted by $m$, are given by $0, \pm 1, \pm2, \cdots$; this is true regardless of whether $S$ is an integer or a half-integer. The classical ground states are of the form ${\vec S}_1 = S (\cos \phi, \sin \phi, 0)$; ${\vec S}_2 = -S (\cos \phi, \sin \phi, 0)$, where the angle $\phi$ can take any value from 0 to $2\pi$. As our reference state, we take the state with $\phi=0$. We define Holstein-Primakoff operators as described in Eqs.~\eqref{eq.hp1}. Switching to canonically conjugate variables $(x_1,p_1)$ and $(x_2,p_2)$ as in Eq.~\eqref{eq.HP_xp}, we find the Hamiltonian
\bea H = - JS^2 + \frac{JS}{2} [ p_1^2 + p_2^2 + x_1^2 + x_2^2 - 2 x_1 x_2 - 2], \label{ham22} \eea
up to order $S$. Defining the linear combinations
\bea P &=& \frac{p_1 + p_2}{\sqrt 2}, ~~~~X = \frac{x_1 + x_2}{\sqrt 2}, 
\non \\
p &=& \frac{p_1 - p_2}{\sqrt 2}, ~~~~x = \frac{x_1 - x_2}{\sqrt 2}, 
\label{xp2} \eea
which form canonically conjugate pairs, we find
\beq H ~=~ JS^2 ~+~ \frac{JS}{2} ~[ P^2 ~+~ p^2 ~+~ 2 x^2 ~-~ 2]. 
\label{ham23} \eeq
In the above expression, the term $P^2$ describes a free particle since
there is no term which depends on $X$. We also have an SHO given by $(JS/2) (p^2 + 2 x^2)$, with frequency $\om = JS \sqrt{2}$. The SHO has a zero point 
energy $\om/2 = JS/\sqrt{2}$. The ground state energy is therefore equal to 
\beq E_0 ~=~ - ~JS^2 ~-~ JS ~\Big( 1 ~-~ \frac{1}{\sqrt{2}} \Big). 
\label{e21} \eeq
The complete energy spectrum is given by
\beq E ~=~ E_0 ~+~ n JS \sqrt{2} ~+~ \frac{JS}{2} r^2, \label{e22} \eeq
where $n=0,1,2,\cdots$ represents the state of the SHO, and $r$ denotes the 
eigenvalue of $P$. To find the possible values of $r$, we note from 
Eq.~\eqref{xp2} that $P = S^z/\sqrt{2S}$
where $S^z = S_1^z + S_2^z$ is a good quantum number. Denoting the
eigenvalues of $S^z$ by $m$, we see that the allowed values of 
$(JS/2) r^2$ are given by $m^2/4$. The energy spectrum is therefore 
\beq E ~=~ E_0 ~+~ n JS \sqrt{2} ~+~ \frac{Jm^2}{4}. \label{e23} \eeq
This is the sum of the spectra of an SHO and a particle on a circle.
Eq.~\eqref{e23} is found to agree well with the numerical results obtained by 
exact diagonalization. Note that the spectrum consists of different branches 
corresponding to $n=0,1,2,\cdots$; these branches are separated from each 
other by a gap equal to $JS \sqrt{2}$. In the lowest branch given by $n=0$, energies are given by $Jm^2/4$, values of order unity or $\mathcal{O}(S^0)$.

The excitations corresponding to $m$ can be though of as describing the 
Goldstone mode which appears
in this system because the classical ground state energy does not depend on 
the angle $\phi$. This mode corresponds to a uniform rotation of both the
spins by the operator given by $P = S^z /\sqrt{2S}$.

\section{Holstein-Primakoff theory for the trimer}
\label{App.HPtrimer}

We now present the HP analysis of the trimer problem. 
We first note that the Hamiltonian commutes with the $z$-component of the total spin, 
$S^z \equiv \sum_{j=1}^3 S_j^z$. The eigenvalues of $S^z$, denoted by $m$, 
take values $0, \pm 1, \pm 2, \cdots$ if $S$ is an integer, and $\pm 1/2, 
\pm 3/2, \pm 5/2, \cdots$ if $S$ is a half-integer.

As described in the main text, the classical ground state space consists of two circles. We consider a reference state with $\phi_1 = 0$, $\phi_2 = - 2\pi /3$ 
and $\phi_3 = - 4\pi /3$. Using HP operators as described in Appendix~\ref{App.HPdimer}, we find the Hamiltonian 
\bea H &=& - ~\frac{3JS^2}{2} ~+~ \frac{S}{2} ~\Big(p_1^2 + p_2^2 + p_3^2 + 
x_1^2 + x_2^2 + x_3^2 \non \\
&&- ~x_1 x_2 - x_2 x_3 - x_3 x_1 - 3\Big), \label{ham32} \eea 
up to order $S$. This is diagonalized by defining Jacobi coordinates,
\bea P &=& \frac{p_1 + p_2 + p_3}{\sqrt 3}, \non \\
p_a &=& \frac{p_1 - p_2}{\sqrt 2}, \non \\
p_b &=& \frac{p_1 + p_2 - 2 p_3}{\sqrt 6}, \label{xp3} \eea
and the corresponding canonically conjugate variables $X, ~x_a$ and $x_b$.
In terms of these variables, we find
\bea H &=& - ~\frac{3 JS^2}{2} ~-~ \frac{3 JS}{2} \non \\
&& + ~\frac{JS}{2} ~[P^2 + p_a^2 + p_b^2 + \frac{3}{2} (x_a^2 + x_b^2)]. 
\label{ham33} \eea
We thus have a free particle described by $P^2$ and two SHO's described
by $(p_a,x_a)$ and $(p_b,x_b)$, which have the same frequency $\om = J S
\sqrt{3/2}$. The ground state energy is therefore given by
\beq E_0 ~=~ - ~\frac{3 JS^2}{2} ~-~ JS \Big(\frac{3}{2} ~-~ 
\sqrt{\frac{3}{2}}\Big). \label{e31} \eeq
The complete energy spectrum is given by
\beq E ~=~ E_0 ~+~ (n_a + n_b) JS \sqrt{\frac{3}{2}} ~+~ \frac{JS}{2} 
r^2, \label{e32} \eeq
where $n_a, n_b = 0, 1, 2, \cdots$ represent the different states of the SHO's, 
and $r$ denotes eigenvalues of $P$.
The operator $P$ is related to the total $S^z$ as 
\beq P ~=~ \frac{S^z}{\sqrt{3S}}. \label{p3} \eeq
Hence, the allowed values of $(JS/2) r^2$ are related to the eigenvalues of 
$S^z = m$ as $Jm^2/6$. The energy spectrum therefore takes the form
\beq E ~=~ E_0 ~+~ (n_a + n_b) JS \sqrt{\frac{3}{2}} ~+~ \frac{Jm^2}{6}.
\label{e33} \eeq
The lowest branch of excitations is given by $n_a = n_b = 0$. In this branch,
the spectrum is that of a particle moving on a circle. The difference between 
the energies of the ground state and first excited state is $(J/6)(1^2 - 0^2)
=J/6$ if $S$ is an integer and $(J/6) [(3/2)^2 - (1/2)^2]=J/3$ if $S$ is a 
half-integer. We again find that these results agree with those obtained 
by exact diagonalization.

The excitations corresponding to $m$ describe the Goldstone mode 
which appears because the classical ground state energy does not depend on 
the angle $\phi_1$. This mode corresponds to a uniform rotation of the 
three spins by the operator in Eq.~\eqref{p3}.

\section{Order by singularity in the classical quadrumer}
\label{App.class_quad}
In the main text, we have discussed the notion of order by singularity for quantum spins. We have presented evidence that the effect survives even in the classical limit, i.e., when $S\rightarrow \infty$. Here, we discuss state selection in a purely classical setting, following the approach of Chalker and Moessner\cite{Moessner1998b} and Chalker\cite{Chalker2011} to the classical quadrumer problem. Working in the limit of low temperatures (weak thermal fluctuations), they evaluate the probability distribution for the angular separation between two particular spins. Surprisingly, when the quadrumer is taken to have $XY$ couplings, this probability distribution shows $\delta$-function-like peaks for relative angles
equal to 0 and $\pi$ (corresponding to collinear configurations as discussed
below). In the light of our analysis in the main text, we revisit this problem with respect to the role of singularities in the CGSS.

\begin{figure}
\includegraphics[width=3in]{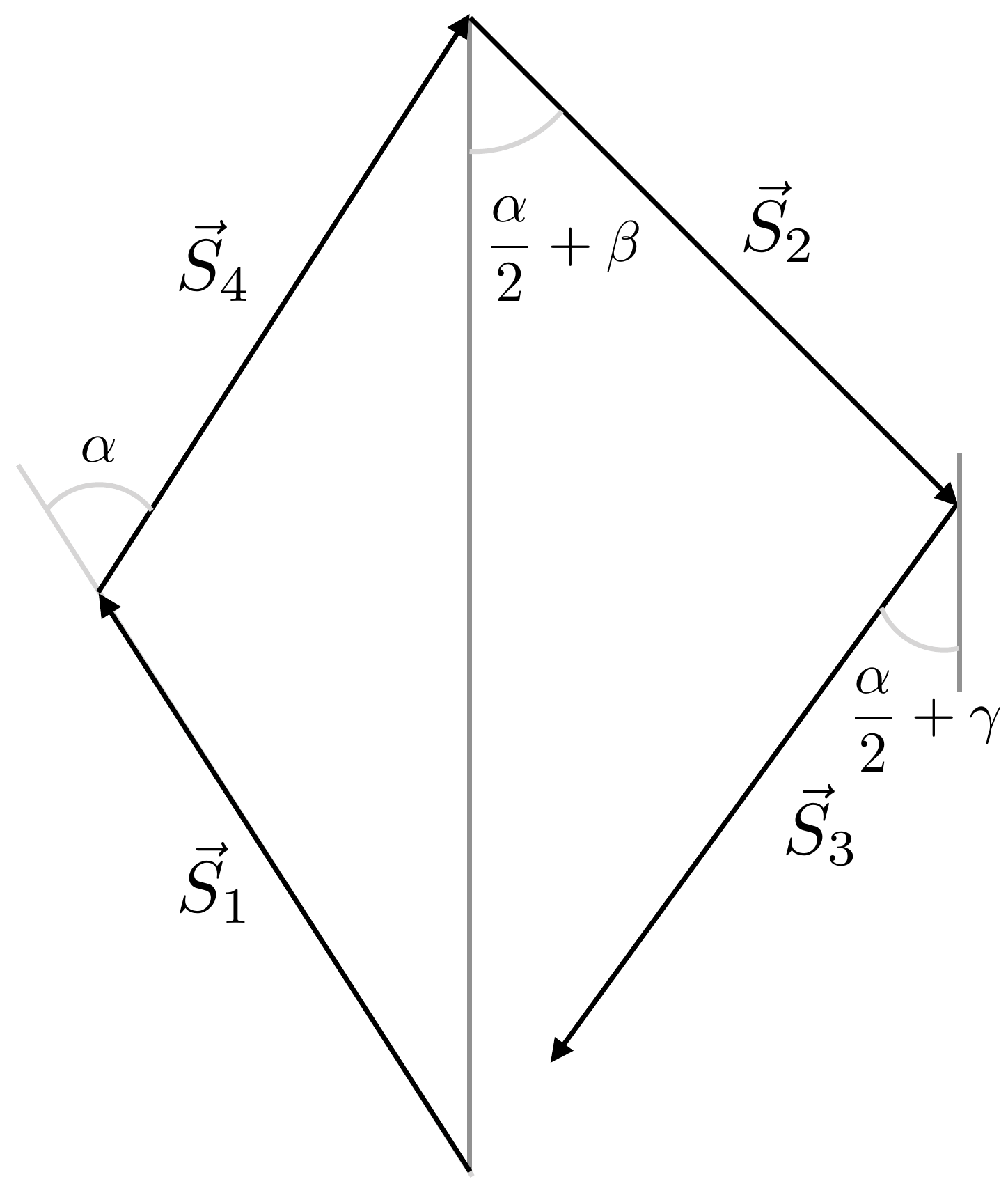}
\caption{Low temperature configuration of the classical $XY$ quadrumer. The angle $\alpha$ can be arbitrary. The configuration is a ground state as long as $\beta =\gamma =0$. At low temperatures, we can assume that $\beta$ and $\gamma$ are small. The figure is reproduced here from Fig. 1.8 in Ref.~\cite{Chalker2011}.}
\label{fig.flucs}
\end{figure}

We consider the classical quadrumer with asymmetry as a tuning parameter. Its Hamiltonian is given by
\begin{equation}
H = \frac{J}{2} \left(\sum_i \mathbf{S}_i\right)^2 + \frac{\lambda J}{2}\left\{ (\mathbf{S}_1+ \mathbf{S}_2)^2+ (\mathbf{S}_3 + \mathbf{S}_4)^2 \right\} .
\end{equation}
Since we are working with purely $XY$ spins, we define 
$(\mathbf{\Sigma})^2 \equiv \Sigma_x \Sigma_x + \Sigma_y \Sigma_y $. 
At low energies, we may restrict our attention to classical ground states and their vicinities. The classical ground states here have $\mathbf{S}_1 = -\mathbf{S}_2$ and $\mathbf{S}_3 = - \mathbf{S}_4$.
We parametrize the spins as (see Fig.~\ref{fig.flucs})
\begin{eqnarray}
\mathbf{S}_1 &=& \left(-\sin(\alpha / 2), \cos(\alpha / 2)\right),\nonumber\\
\mathbf{S}_4 &=& \left(\sin(\alpha / 2), \cos(\alpha / 2)\right),\nonumber\\
\mathbf{S}_2 &=& \left(\sin(\alpha / 2 + \beta), -\cos(\alpha / 2 + \beta)\right),\nonumber\\
\mathbf{S}_3 &=& \left(-\sin(\alpha / 2 + \gamma ), -\cos(\alpha / 2 + \gamma) \right)
\end{eqnarray}
Here, $\beta$ and $\gamma$ represent small fluctuations that take us away from the ground state space. The energy takes the form
\begin{eqnarray}
H &=&\frac{JS^2}{2}\Big((1 + \lambda)\beta ^2 + (1+\lambda)\gamma^2 - 2\beta\gamma\cos(\alpha)\Big)\nonumber\\
&=&\frac{JS^2}{2}(1+\lambda)\left[\left(\beta - \frac{\cos(\alpha)}{1+\lambda}\gamma\right)^2 + \left(1 - \frac{\cos^2(\alpha)}{(1+\lambda)^2}\right)\gamma^2\right].\nonumber
\end{eqnarray}
In order to integrate out fluctuations, we make the following variable change:
\begin{eqnarray}
\delta_1 & = & \beta - \frac{\cos(\alpha)}{1+\lambda}\gamma\nonumber\\
\delta_2 & = & \gamma,
\end{eqnarray}
a transformation for which the Jacobian is unity. In terms of the new variables, the Hamiltonian is
\begin{equation}
H = \frac{JS^2}{2}(1+\lambda)\left[\delta_1^2 + \left(1 - \frac{\cos^2(\alpha)}{(1+\lambda)^2}\right)\delta_2^2\right]
\end{equation}
We now integrate out the $\delta$'s. Setting $k_B$ to unity, we obtain the probability distribution for $\alpha$, the angle between spins $S_1$ and $S_4$, as 
\begin{eqnarray}
P(\alpha) &=& \int_{\delta_1}\int_{\delta_2} \exp(-H/T)\nonumber\\
&\sim& T\times\sqrt{\frac{1}{(1 + \lambda)^2 - \cos^2(\alpha)}}
\label{probability}
\end{eqnarray}
We see that the probability is finite for all $\alpha$ as long as $\lambda > 0$. When $\lambda =0 $, we recover the results of Chalker and Moessner, with non-integrable divergences at $\alpha = 0,\pi$.
An infinitesimal asymmetry suffices to remove the sharp selection associated with divergences. As discussed in the main text, a small asymmetry term also removes singularities in the CGSS. This suggests that singularities could play a role in sharp state selection.

\section{Tangent space on the non-manifold CGSS}
\label{App.tangent}
In the main text, we have shown that the CGSS of the symmetric quadrumer is not a manifold due to the presence of singular lines. Here, we discuss its tangent space to explicitly bring out the non-manifold character. 

We denote the position of the $i^{\mathrm{th}}$ spin 
as $\mathbf{S}_i = \left(S_{ix},S_{iy}, S_{iz}\right)$, with $i=1,2,3,4$. A generic state of a quadrumer is represented by a twelve-dimensional vector $\left(\mathbf{S}_1,\mathbf{S}_2, \mathbf{S}_3, \mathbf{S}_4\right)$. In order to represent a genuine spin configuration, we must have 
\begin{equation} S_{ix}^2 + S_{iy}^2 + S_{iz}^2 = 1 ~~~{\rm for}~~~ i=1,2,3,4.
\end{equation}
(We have taken the spin length to be unity). This amounts to four constraints on the twelve-dimensional vector. 

As discussed in the main text, the minimum energy configurations must have (a) spins lying in the $x-y$ plane and (b) zero total spin. This is equivalent to the following six constraints,
\begin{eqnarray}
S_{iz} &=& 0, ~~~{\rm for}~~~ i=1,2,3,4, \non \\ 
\sum_{i=1}^4 S_{ix} &=& \sum_{i=1}^4 S_{iy} =0. \end{eqnarray}
Ground state configurations must necessarily have two pairs of anti-aligned spins. We consider a generic ground state with $\vec{S}_1 = -\vec{S}_3$ and $\vec{S}_2 = -\vec{S}_4$. In the twelve-dimensional configuration space, this corresponds to 
\begin{eqnarray}
\nonumber \vec{P} &=& \Big(\cos\phi_1, \sin\phi_1, 0, 
 \cos\phi_2, \sin\phi_2, 0,\\
&~& -\cos\phi_1, -\sin\phi_1, 0, -\cos\phi_2, -\sin\phi_2, 0\Big).
\label{eq.Sigma}
\end{eqnarray}
We have chosen $\vec{S}_1$ and $\vec{S}_2$ to make angles $\phi_1$ and $\phi_2 $ with the $\hat x$-axis respectively.

We now analyze all the independent fluctuations about this ground state. The fluctuation modes fall into the following three categories:\\
\vspace{0.1 in}
\textit{1. Soft deformations:} These modes preserve the spin lengths as well as the ground state conditions. We enumerate them as:\\
\begin{eqnarray}
\label{eq.gen_soft}
\hat{\sigma}_1 &=& \frac{1}{2}\Big(\sin\phi_1, -\cos\phi_1, 0 , \sin\phi_2, -\cos\phi_2, 0,\nonumber\\
&~& -\sin\phi_1, \cos\phi_1, 0, -\sin\phi_2, \cos\phi_2, 0 \Big),\nonumber\\
\nonumber \hat{\sigma}_2 &=& \frac{1}{2}\Big(\sin\phi_1, -\cos\phi_1, 0, -\sin\phi_2, \cos\phi_2, 0, \\
&~& -\sin\phi_1, \cos\phi_1, 0, \sin\phi_2, -\cos\phi_2, 0 \Big).
\label{eq.sigma_vecs}
\end{eqnarray}
These are orthogonal to each other as well as to the reference state $\vec{P}$. Here, orthogonality is defined using the dot product in the twelve-dimensional embedding space. Physically, the mode $\hat{\sigma}_1 $ represents overall rotation in the $x-y$ plane. The mode $\hat{\sigma}_2$ describes a scissor-like deformation between two rods, one consisting of $(\vec{S}_1, \vec{S}_3)$ and the other composed of $(\vec{S}_2,\vec{S}_4)$. 
The soft nature of these modes can be seen by constructing $\vec{P}_{s,\delta} = \vec{P} + \sum_{i = 1}^2 \delta_i \hat{\sigma}_i $, where $\delta_i \ll 1$ represent small deviations from the reference state. To linear order in the
$\delta_i$'s, $\vec{P}_{s,\delta}$ satisfies the spin length constraints. 
In addition, it readily satisfies the ground state conditions.
Thus, $\vec{P}_{s,\delta}$ represents the local neighborhood of a point on the ground state space, i.e., it represents the tangent space to the CGSS at $\vec{P}$.

\textit{2. Hard deformations:}
We next consider modes that preserve the spin lengths but not the ground state energy. We enumerate them as: 
\begin{eqnarray}
\label{eq.gen_hard}
\nonumber \hat{h}_1 &=& \frac{1}{2}\Big(\sin\phi_1, -\cos\phi_1, 0 , -\sin\phi_2, \cos\phi_2, 0, \\
&~& \sin\phi_1, -\cos\phi_1, 0, -\sin\phi_2, \cos\phi_2, 0 \Big),\nonumber\\
\nonumber \hat{h}_2 &=&\frac{1}{2}\Big(-\sin\phi_1, \cos\phi_1, 0 , -\sin\phi_2, \cos\phi_2, 0, \\
&~& -\sin\phi_1, \cos\phi_1, 0, -\sin\phi_2, \cos\phi_2, 0 \Big),\nonumber\\
\hat{h}_3 &=& \left( 0,0,1,0,0,0,0,0,0,0,0,0\right),\nonumber\\
\hat{h}_4 &=& \left( 0,0,0,0,0,1,0,0,0,0,0,0\right),\nonumber\\ 
\hat{h}_5 &=& \left( 0,0,0,0,0,0,0,0,1,0,0,0\right),\nonumber\\
\hat{h}_6 &=& \left( 0,0,0,0,0,0,0,0,0,0,0,1\right).
\end{eqnarray}
These are orthogonal to each other, to the soft modes as well as to the reference state $\vec{P}$. We consider $\vec{P}_{h,\eta} = \vec{P} + \sum_{i = 1}^6 \eta_i \hat{h}_i $, with the $\eta_i$'s representing small amplitudes. We find that $\vec{P}_{h,\eta}$ preserves spin lengths to linear order in the $\eta_i$'s. However, it violates the ground state conditions with energy increasing quadratically in the $\eta_i$'s. We conclude that these modes are physically allowed fluctuations that take the system out of the ground state space. \\
\vspace{0.1 in }
\textit{3. Unphysical deformations}:
As the embedding space is twelve-dimensional, we must have twelve fluctuation modes about any configuration. We have enumerated two soft modes and six hard modes above. The remaining four modes represent unphysical deformations that violate the spin length constraints. They are
\begin{eqnarray}
\label{eq.gen_unphy}
\hat{u}_1 &=& \left( \cos\phi_1,\sin\phi_1,0,0,0,0,0,0,0,0,0,0\right),\nonumber\\
\hat{u}_2 &=& \left( 0,0,0,\cos\phi_2,\sin\phi_2,0,0,0,0,0,0,0\right),\nonumber\\ 
\hat{u}_3 &=& \left( 0,0,0,0,0,0, -\cos\phi_1,-\sin\phi_1,0,0,0,0\right),\nonumber\\
\hat{u}_4 &=& \left( 0,0,0,0,0,0,0,0,0,-\cos\phi_2,-\sin\phi_2,0\right).
\end{eqnarray}
These modes are orthogonal to each other as well as to the soft and hard modes. The spin lengths are not preserved to linear order in these fluctuations -- making them unphysical. \\
 
The above analysis shows that, about a generic ground state, there are two soft modes, six hard modes and four unphysical modes. The tangent space to the CGSS is two-dimensional, spanned by the two soft deformations. However, a different picture emerges in the vicinity of collinear ground states. To see this, we consider $\phi_1 = \phi_2 = 0$ in Eq.~\eqref{eq.Sigma}. This corresponds to $\vec{S}_1 = \vec{S}_2 = -\vec{S}_3 = -\vec{S}_4 = (1,0,0)$. The $\hat{\sigma}$ modes given in Eq.~\eqref{eq.sigma_vecs} retain their soft mode character. However, among the $\hat{h}$ modes, $\hat{h}_1 = \frac{1}{2}\left(0,-1,0,0,1,0,0,-1,0,0,1,0\right) $ changes its character. It `softens' as it no longer violates the ground state conditions. 
This is reminiscent of excitations in the triangular XY antiferromagnet wherein a hard mode softens at a critical magnetic field\cite{Chubokov1991}. Here, in the vicinity of collinear states, we have three soft modes, five hard modes and four unphysical modes. The tangent space is now three-dimensional, represented by $\vec{P}_{s,coll.,\mu,\eta} = \vec{P}(\phi_1 = \phi_2=0) + \mu_1 \hat{\sigma}_1 + \mu_2 \hat{\sigma}_2 + \eta_1 \hat{h}_1$. 

The tangent space cannot be defined smoothly on the CGSS due to differing dimensionalities. This marks the CGSS as a non-manifold space. Our tight-binding model is designed to take this into account.
To see this, we refer to the three-torus geometry of the CGSS as discussed in the main text. Each torus is discretized with two local directions as shown in Fig.~\ref{fig.mesh_metric}, corresponding to a global rotation and a `scissor' deformation. These are precisely the deformations induced by $\hat{\sigma}_1$ and $\hat{\sigma}_2$ above.
At the singular lines, the tight-binding particle is allowed to move in a third direction, corresponding to motion away from one torus onto a second torus. In the example with $\phi_1 = \phi_2 = 0$ discussed above, this corresponds precisely to the softened $\hat{h}_1$ mode. To see this, we examine $\vec{P}(\phi_1 = \phi_2=0) + 2\eta_1 \hat{h}_1 = (1,-\eta_1,0,1,\eta_1,0,-1,-\eta_1,0,-1,\eta_1,0)$. This deformed state has $(\vec{S}_1 = -\vec{S}_4;~ \vec{S}_2 = -\vec{S}_3)$. Note that this configuration of alignments is different from $(\vec{S}_1 = -\vec{S}_3;~\vec{S}_2 = -\vec{S}_4)$ that was assumed in Eq.~\eqref{eq.Sigma}. By our definition, these correspond to two different tori. In summary, the softened mode at collinear lines corresponds to motion from one torus to another -- a possibility that is explicitly allowed in our tight-binding model. 

\bibliographystyle{apsrev4-1}
\bibliography{quadrumer}

\end{document}